\documentclass[fleqn,usenatbib]{mnras}

\usepackage{newtxtext,newtxmath}

\usepackage[T1]{fontenc}

\DeclareRobustCommand{\VAN}[3]{#2}
\let\VANthebibliography\thebibliography
\def\thebibliography{\DeclareRobustCommand{\VAN}[3]{##3}\VANthebibliography}

\usepackage{graphicx}	
\usepackage{amsmath}	
\usepackage[]{units}

\usepackage{soul}

\newcommand{\mspc}{M_\odot\,\mathrm{pc}^{-2}}

\title[Great Balls of FIRE]{Great Balls of FIRE I: The formation of star clusters across cosmic time in a Milky Way-mass galaxy}
\author[Grudi\'{c} et al.]
{Michael Y. Grudi\'{c}$^{1}$\thanks{mgrudic@carnegiescience.edu}\thanks{NASA Hubble Fellow},
Zachary Hafen$^{2}$,
Carl L.~Rodriguez$^{3}$,
D\'{a}vid Guszejnov$^{4}$,\newauthor
Astrid Lamberts$^{5,6}$,
Andrew Wetzel$^{7}$,
Michael Boylan-Kolchin$^{4}$,\newauthor
and Claude-Andr\'{e} Faucher-Gigu\`{e}re$^{8}$
\\
$^{1}${Carnegie Observatories, 813 Santa Barbara St, Pasadena, CA 91101, USA}\\
$^{2}$Center for Cosmology, Department of Physics and Astronomy, University of California, Irvine, CA, 92697, USA\\
$^{3}${McWilliams Center for Cosmology and Department of Physics, Carnegie Mellon University, Pittsburgh, PA 15213, USA}\\
$^{4}$Department of Astronomy, The University of Texas at Austin, TX 78712, USA \\
$^{5}$Laboratoire Lagrange, Observatoire de la C\^{o}te d'Azur, Universit\'{e} C\^{o}te d'Azur, CNRS, CS 34229, F-06304 NICE Cedex 4, France\\
$^{6}${Laboratoire Artemis, Observatoire de la C\^{o}te d'Azur, Universit\'{e} C\^{o}te d'Azur, CNRS, CS 34229, F-06304 Nice Cedex 4, France}\\
$^{7}${Department of Physics and Astronomy, University of California, Davis, CA 95616, USA} \\
$^{8}${CIERA and Department of Physics and Astronomy, Northwestern University, 1800 Sherman Ave, Evanston, IL 60201, USA}\\
}

\date{Accepted XXX. Received YYY; in original form ZZZ}

\defcitealias{grudic:2020.cluster.formation}{G21}
\defcitealias{kruijssen:2012.cluster.formation.efficiency}{K12}

\pubyear{2022}
\begin{document}
\label{firstpage}
\pagerange{\pageref{firstpage}--\pageref{lastpage}}
\maketitle

\begin{abstract}
The properties of young star clusters formed within a galaxy are thought to vary in different interstellar medium (ISM) conditions, but the details of this mapping from galactic to cluster scales are poorly understood due to the large dynamic range involved in galaxy and star cluster formation. We introduce a new method for modeling cluster formation in galaxy simulations: mapping giant molecular clouds (GMCs) formed self-consistently in a FIRE-2 MHD galaxy simulation onto a cluster population according to a GMC-scale cluster formation model calibrated to higher-resolution simulations, obtaining detailed properties of the galaxy's star clusters in mass, metallicity, space, and time. We find $\sim 10\%$ of all stars formed in the galaxy originate in gravitationally-bound clusters overall, and this fraction increases in regions with elevated $\Sigma_{\rm gas}$ and $\Sigma_{\rm SFR}$, because such regions host denser GMCs with higher star formation efficiency. These quantities vary systematically over the history of the galaxy, driving variations in cluster formation. The mass function of bound clusters varies -- no single Schechter-like or power-law distribution applies at all times. In the most extreme episodes, clusters as massive as $7\times 10^6 M_\odot$ form in massive, dense clouds with high star formation efficiency. The initial mass-radius relation of young star clusters is consistent with an environmentally-dependent 3D density that increases with $\Sigma_{\rm gas}$ and $\Sigma_{\rm SFR}$. The model does not reproduce the age and metallicity statistics of old ($>11\rm Gyr$) globular clusters found in the Milky Way, possibly because it forms stars more slowly at $z>3$.
\end{abstract}

\begin{keywords}
galaxies: star formation -- galaxies: star clusters: general -- open clusters and associations: general -- ISM: clouds -- globular clusters: general
\end{keywords}



\section{Introduction}

Stars can form either as members of unbound associations of stars that will disperse into the host galaxy, or as members of gravitationally-bound star clusters (hereafter ``star clusters" or ``clusters") that can persist for significantly longer \citep{gouliermis:2018.review,adamo:2020.clusters.review, ward.kruijssen:2018}. The persistence of bound clusters is interesting because they are lasting, coherent relics of star formation events whose properties are thought to bear some imprint of their natal environment. Their evolution, and eventual demise, are shaped by both stellar dynamics and the galactic gravitational landscape \citep{fall.zhang:2001.cluster.evol,gieles:2006.cluster.evolution}, so a galaxy's star cluster population contains a record of both its formation and its dynamical history.

Deciphering this record requires an understanding of the evolution of the star-forming interstellar medium (ISM) within galaxies, {\it and} how these changing conditions map onto to the properties of young star clusters. Observations point to several such connections. The fraction of star formation in bound clusters (cluster formation efficiency, CFE, commonly denoted $\Gamma$, \citealt{bastian:2008.cfe}) has been found to vary, both from one galaxy to another and between different regions of a given galaxy (\citealt{goddard:2010.cfe,cook:2012.dwarf.clusters, adamo:2015.m83.clusters,johnson:2016.cluster.formation.efficiency, chandar:2017.cfe,ginsburg:2018.cfe}; see \citealt{krumholz:2018.star.cluster.review, adamo:2020.clusters.review} for review). The {\it initial mass function} (hereafter simply ``mass function") of young clusters may also vary: various works have reported evidence of a high-mass truncation at a certain characteristic cutoff mass scale \citep{gieles:2006.m51.massfunc.cutoff,adamo:2015.m83.clusters, johnson:2017.m31.massfunction,messa:2018.m51.2,2022arXiv220104161W}, with reported values ranging from $\sim 10^4-10^6 M_\odot$, generally between regions of low and high star formation intensity $\Sigma_{\rm SFR}$ \footnote{See however \citet{mok:2019.clusters}, who found the significance of various reported mass function cutoff masses in the literature to be marginal, and \citet{2022arXiv220104161W} who explored how uncertainties in the few greatest cluster masses propagate into the uncertainty of the cutoff mass.}. And recently, several works have noted possible variations in the {\it mass-radius} relation of young star clusters, with more intensely star-forming environments hosting more compact clusters for a given mass \citep{krumholz:2018.star.cluster.review,choksi:2019.cluster.mass.radius,grudic:2020.cluster.formation,brown:2021.cluster.mass.radius}. In all, it is clear that there is an intimate connection between star-forming environment and the properties of young star clusters.



To understand the physical processes driving variations in cluster properties across cosmic time, we require a model that couples the full cosmological context of galaxy formation to the formation and evolution of individual star clusters in. Resolution requirements make this presently impossible to do in direct calculations that track the formation of individual stars \citep[e.g.][]{bate2003,krumholz:2011.rhd.starcluster.sim,haugbolle2018,starforge.methods}, so a number of approximate frameworks have been devised to model the formation and evolution of star clusters in galaxy simulations, accounting for various subsets of the relevant physics either self-consistently or with sub-resolution models. Simulations using a sub-grid ISM model \citep[e.g.][]{springel:2003.subgrid.ism} do not follow the formation of individual giant molecular clouds (GMCs), so the population of GMCs are modeled according to the available bulk ISM properties on $\sim \rm kpc$ scales (e.g. density, pressure, and metallicity), and the mapping from GMCs to stars and bound clusters in turn via semi-analytic models \citep{kruijssen:2011.mosaics,pfeffer:2018.emosaics}. \citet{li:2017.cluster.formation} performed cosmological galaxy simulations with sufficient resolution to resolve some individual GMCs, and modeled cluster formation in them as unresolved accreting sink particles with sub-grid feedback injection \citep{agertz:2013.new.stellar.fb.model,semenov:2016.sgs.turb}, internal structure, and dynamical evolution \citep{gnedin:2014.cluster.evol}. Other galaxy simulations resolving as fine as $\sim 1 \rm pc$ scales have modeled clusters as bound collections of softened subcluster particles \citep{kim:2017.globular.clusters,lahen_2019_cluster_formation, ma_2020_fire_cluster_formation}, using sub-grid prescriptions for star formation and feedback coupled on the relevant resolved scale.

Each of the approaches listed above has different advantages and potential pitfalls, but all rely upon somewhat uncertain prescriptions for unresolved star formation and stellar feedback, which have not been explicitly validated due to the difficulty of simulating star-forming GMCs self-consistently. Lacking a definitive numerical model for cosmological cluster formation and evolution, it is worthwhile to consider alternative approaches to treating cosmological star cluster formation in galaxies, especially ones that can be applied to {\it existing} galaxy simulations without modification.

In this work we introduce a new post-processing technique for modeling star cluster formation in existing cosmological simulations that resolve the multi-phase interstellar medium:
we map the properties of GMCs formed in a FIRE-2 cosmological zoom-in simulation \citep{wetzel:2016.latte, hopkins:2020.whatabout, guszejnov_GMC_cosmic_evol} onto a model star cluster population via a statistical model calibrated to high-resolution ($\sim 0.1$ pc) cluster formation simulations with stellar feedback~(\citealt{grudic:2020.cluster.formation}, hereafter \citetalias{grudic:2020.cluster.formation}). This produces definite predictions for the detailed formation efficiency, masses, formation times, metallicities, and initial sizes of star clusters, which can be compared with observed young star cluster catalogues and used as the initial conditions for detailed dynamical treatments of star cluster evolution (Rodriguez et al., in prep.).

This paper is structured as follows. In \S\ref{section:methods} we describe our GMC and star cluster population modeling technique based upon coupling the results of \citet{guszejnov_GMC_cosmic_evol} and \citetalias{grudic:2020.cluster.formation}, and describe the Milky Way-mass galaxy model that we use as a case study. In \S\ref{section:results} we present the results of our model, showing how the efficiency of bound cluster formation, the cluster initial mass function, and cluster size statistics vary across cosmic time according to the evolving ISM conditions in the galaxy. We also examine the properties of the clusters in age-metallicity space, and compare and contrast those statistics with those those of Milky Way globular clusters to comment on the viability of the simulation as a model for globular cluster formation. In \S\ref{section:discussion} we discuss various implications of our results and compare and contrast our findings with previous treatments of cosmological star cluster formation. In \S\ref{section:conclusions} we summarize our key conclusions about the connection between galactic environment and star cluster formation.

\section{Methods}
\begin{figure*}
\includegraphics[width=\textwidth]{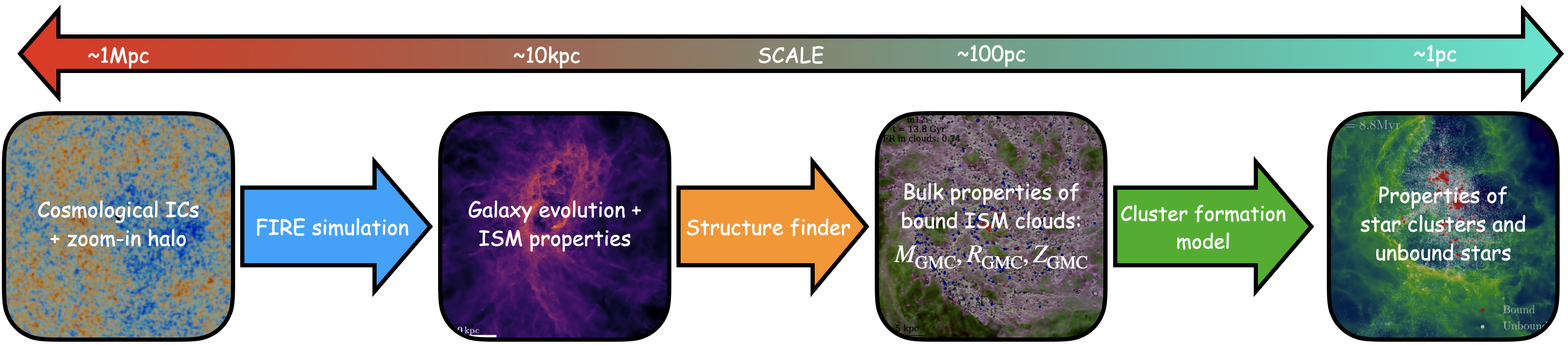}
\caption{Diagram of our procedure for modeling the star cluster population of a simulated galaxy across cosmic time, described in full in \S\ref{section:methods}. Starting with cosmological initial conditions and a choice of zoom-in halo, we simulate the cosmological evolution of the halo to $z=0$ with FIRE \citep{hopkins:2013.fire,wetzel:2016.latte,fire2}, run a structure finder to determine the bulk properties of bound clouds in the ISM \citep{guszejnov_GMC_cosmic_evol}, and plug these cloud properties into a model that predicteds detailed star cluster properties, calibrated to high-resolution GMC simulations \citep{grudic:2020.cluster.formation}. }
\label{fig:flowchart}
\end{figure*}

Our model of galactic star cluster formation has three steps: the FIRE-2 cosmological zoom-in galaxy simulation itself, the extraction of cloud properties from the simulation data, and the mapping of cloud properties onto star cluster properties via the model derived from the  \citetalias{grudic:2020.cluster.formation} GMC simulations. We visualize the procedure in Figure \ref{fig:flowchart} and describe each step in turn below.

\label{section:methods}
\subsection{FIRE-2 simulation}
Here we study the formation of a Milky Way-mass disk galaxy formed in a cosmological zoom-in simulation of the halo model {\texttt m12i} simulated as part of the FIRE-2 simulation suite \citep{fire2} with the {\small GIZMO} code \citep{hopkins:gizmo}. This galaxy simulation accounts for a wide range of relevant cooling mechanisms down to 10 K via detailed fits and tables \citep{fire2}, stellar radiative feedback including radiation pressure, photoionization, and photoelectric heating \citep{FIRE_RT}, OB/AGB stellar winds, and type Ia and II supernovae \citep{FIRE_SNE}, with rates derived from a standard simple stellar population model \citep{starburst99} assuming a \citet{kroupa:imf} stellar initial mass function. The simulation also accounts for magnetic fields using the quasi-Lagrangian, Meshless Finite Mass (MFM) magnetohydrodynamics solver \citep{hopkins:gizmo.mhd}, anisotropic Spitzer-Braginskii conduction and viscosity, and sub-grid metal diffusion from unresolved turbulence \citep{hopkins:2017.diffusion, su:2016.feedback.first, hopkins:2020.whatabout}. 

At $z=0$ the simulated galaxy has a stellar mass of $M_{\rm \star} = 6.7\times 10^{10} M_\odot$ and a halo mass of $M_{\rm 200} = 1.2 \times 10^{12} M_\odot$, similar to inferred present-day mass measurements of the Milky Way \citep{blandhawthorn:2016.milkyway}. See \citet{2020ApJS..246....6S} for various detailed comparisons of non-MHD version of this simulation with the Milky Way, and \citet{2020MNRAS.498.3664G} for a detailed analysis of the phase structure and dynamics of its interstellar medium. We selected the version of the simulation with MHD, conduction, and viscosity as a more physically-complete model, note that the incremental effects of such processes upon star formation and galaxy evolution in this simulation have been shown to be modest \citep{su:2016.feedback.first,hopkins:2020.whatabout}.

\subsection{Cloud catalogue}
\label{sec:catalogue}
The galaxy simulation has a baryonic mass resolution of $7070$ $ M_\odot$ and accounts for detailed multi-phase ISM physics, allowing it to resolve the bulk properties of massive ($\gtrsim 10^5 M_\odot$) giant molecular clouds (GMCs). The GMCs in the simulation assemble, form stars, and disperse in the simulation self-consistently over a typical timescale on the order of $10 \rm Myr$ \citep{hopkins:fb.ism.prop,benincasa:2020.fire.gmcs}. In \citet{guszejnov_GMC_cosmic_evol} we used {\small CloudPhinder}\footnote{\url{http://www.github.com/mikegrudic/CloudPhinder}}, an algorithm similar to {\small SUBFIND} \citep{springel:2001.subfind}, to identify the population of {\it self-gravitating} gas structures in this simulation: specifically, our algorithm identifies the population of 3D iso-density contours that enclose material with virial parameter $\alpha_{\rm vir} <2$ \citep{bertoldi.mckee}.

\citet{guszejnov_GMC_cosmic_evol} found these objects to have surface densities, size-linewidth relations, and maximum masses that resemble GMCs found in nearby galaxies \citep[e.g.][]{larson:gmc.scalings, bolatto:2008.gmc.properties, colombo:2014.m51.gmcs, freeman:2017.m83.gmcs, faesi:2018.ngc300.gmcs}. However, the internal structure and dynamics of the clouds remain largely unresolved at our mass resolution, so we do not generally expect the star clusters that form in clouds to have reliable properties uncontaminated by numerical effects. Hence, we synthesize the cluster population in post-processing, using the properties of the self-gravitating clouds catalogued in \citet{guszejnov_GMC_cosmic_evol} as inputs to our star cluster formation model. We adopt a minimum mass cut of $2\times 10^5 M_\odot$, or $\sim 30$ times the mass resolution.


\subsection{Cluster formation model}
\label{section:model}
To determine the properties of clusters formed in the GMCs, we adopt the cluster formation model introduced in \citetalias{grudic:2020.cluster.formation}. In that study we performed a large suite of MHD star cluster-forming GMC simulations including stellar feedback, finding that quantities such as star formation efficiency, the fraction of star formation in bound clusters, and individual star cluster masses do depend sensitively upon the macroscopic properties of the parent GMC such as mass $M_{\rm GMC}$ and size $R_{\rm GMC}$, but may also vary strongly from one GMC to another even if these quantities are held fixed, due to variations in the details of the initial turbulent flow. This led us to develop a {\it statistical} model that reproduces the statistical results (e.g. cluster mass functions and size distributions) of the ensemble of simulation results over many different initial realizations of turbulence. By modeling star cluster formation in this way, we arrived at a model that could reproduce the CFE and young star cluster mass functions observed in M83 \citep{adamo:2015.m83.clusters} fairly well, if the observed properties of GMCs in those respective galactic regions were taken as inputs \citep{freeman:2017.m83.gmcs}.

We briefly summarize the procedure of the model here. Given the the mass $M_{\rm GMC}$, size $R_{\rm GMC}$, and metallicity $Z_{\rm GMC}$ of a cloud, the calculation proceeds as follows. First, we determine the total stellar mass formed in the cloud:

\begin{equation}
    M_{\rm \star} = \epsilon_{\rm int} M_{\rm GMC},
\end{equation}
where $\epsilon_{\rm int}$ is the integrated star formation efficiency, which depends upon the GMC surface density $\Sigma_{\rm GMC} =M_{\rm GMC}/\uppi R_{\rm GMC}^2$ as:

\begin{equation}
    \epsilon_{\rm int} = \left(\epsilon_{\rm max}^{-1} + \left(\frac{\Sigma_{\rm GMC}}{3200\,M_\odot \rm pc^{-2}}\right)^{-1} \right)^{-1} \approx \frac{\Sigma_{\rm GMC}}{3200\,M_\odot\,\rm pc^{-2}},
    \label{eq:sfe}
\end{equation}
where $\epsilon_{\rm max}=0.7$ and the latter approximation holds when $\Sigma_{\rm GMC} \lesssim 3000 \,M_\odot \rm pc^{-2}$.

With the total stellar mass known, we then determine the fraction of stars locked into bound clusters $f_{\rm bound}$. \citetalias{grudic:2020.cluster.formation} found $f_{\rm bound}$ to vary as a function of $\Sigma_{\rm GMC}$ and $Z_{\rm GMC}$, but the significant scatter from one realization to another requires a probablistic model. Specifically, we let
\begin{equation}
    f_{\rm bound} = \left(1 + \left(\frac{\Sigma_{\rm GMC}}{\mathrm{e}^{\delta} \Sigma_{\rm bound}  }\right)^{n}\right)^{-1},
    \label{eq:fbound}
\end{equation}
where the random variable $\delta $ is sampled from a log-normal distribution with mean $\mu = 0$ and width $\sigma = 0.3 \rm dex$, and the metallicity-dependent parameters $\Sigma_{\rm bound}$ and $n$ are 
\begin{equation}
    \Sigma_{\rm bound} = \left(30 \log Z_{\rm GMC} + 390 \right) \mspc
\end{equation}
and 
\begin{equation}
    n = -0.3\log Z_{\rm GMC} - 2,
\end{equation}
where $Z_{\rm GMC}$ is the GMC metallicity in solar units. We found this prescription to reproduce the scaling of $f_{\rm bound}$ with GMC properties, and its intrinsic scatter across different realizations of a given set of bulk properties. 

The total bound stellar mass is then $M_{\rm bound} = f_{\rm bound} \epsilon_{\rm int} M_{\rm GMC}$. The simulations generally found significant {\it multiplicity} of clusters formed in a single parent GMC, so we distribute the bound mass among the individual clusters by sampling from a GMC-level mass distribution, given in \citetalias{grudic:2020.cluster.formation} Eq. 11. Then, given the list of cluster masses, we determine their half-mass radii $r_{\rm h}$ by sampling a GMC-level size-mass relation:
\begin{equation}
    r_{\rm h} = \unit[3]{pc}\left(\frac{M_{\rm GMC}}{\unit[10^6]{M_\odot}}\right)^\frac{1}{5}\left( \frac{\Sigma_{\rm GMC}}{\unit[100]{\mspc}} \right)^{-1} \left(\frac{Z_{\rm GMC}}{Z_\odot}\right)^\frac{1}{10} \left(\frac{M_{\rm cl}}{\unit[10^4]{M_\odot}}\right)^\frac{1}{3},
    \label{eq:sizefit}
\end{equation}

with an intrinsic log-normal scatter of  $\pm \unit[0.4]{dex}$ in radius. Lastly, although we do not require the detailed density profile for the present work, it is eventually required to model the dynamical evolution and observational characteristics of the clusters. We assume the clusters initially have a \citet{Elson:1987.ymc.profile} density profile and sample the density profile slope $\gamma$ from a universal distribution consistent with observations \citet{grudic:2017}.



For the purposes of the present analysis, we apply a lower mass cut of $10^3 M_\odot$ to the cluster catalogue, similar to the completeness limits of extragalactic cluster catalogues \citep{adamo:2015.m83.clusters,messa:2018.m51.2,johnson:2017.m31.massfunction}. Note that our model will have its own incompleteness function due to our lower GMC mass cutoff of $2 \times 10^5 M_\odot$ -- how this maps onto a cluster mass scale will depend upon the detailed SFE and CFE statistics of the cloud sample.

\subsubsection{Sampling procedure}
\label{sec:sampling}
The clouds selected by the cloud-finding algorithm are not a complete census of all clouds to ever form stars within the model galaxy. The simulation has 601 snapshots, which can be spaced as far apart as $\sim \rm 26$ Myr, likely significantly longer than the life-time of all but the most massive GMCs, which, at least in FIRE and similar simulations \citep{hopkins:fb.ism.prop,benincasa:2020.fire.gmcs, li:2020.smuggle.gmcs}, and observations \citep{chevance:2020.gmcs}, is generally on the order of the cloud freefall time, $\sim 3-10 \,\rm Myr$. 
Therefore, clouds in the simulation typically form and disperse between snapshots (which are typically $\sim 22 \rm Myr$ apart), preventing them from being found by our structure finder. Furthermore, we have found in previous high-resolution GMC simulations \citep{grudic:2018.gmc.sfe} that a significant fraction of star formation within a cloud can happen when it is already in a super-virial state due to feedback from the first massive stars that formed in it -- under such conditions, the cloud would not be identified by our algorithm, even if it is present in the snapshot. Therefore, a simple 1-to-1 mapping of catalogued clouds to stellar populations will tend to underestimate the total stellar mass in our setup.

To address this issue, we adopt the following sampling procedure to synthesize the cluster population while matching the simulated star formation history, from each snapshot:
\begin{enumerate}
    \item Measure the total galactic stellar mass $\Delta M_{\rm \star}^{\rm gal}$ {\it actually formed} in the simulation in the time between snapshots $i$ and $i+1$.
    \item Sample from the catalogue of clouds found in snapshot $i$ randomly until the total stellar mass formed by the cloud sample according to the \citetalias{grudic:2020.cluster.formation} model exceeds $\Delta M_{\rm \star}^{\rm gal}$. 
\end{enumerate} 

In this way, we use the bound clouds as statistical tracers of the full population of progenitor clouds, and recover a model that accounts for the entire stellar mass of the galaxy. Note that while we are requiring 100\% of the stellar mass to be formed in the bound clouds, only a fraction of that mass will be in bound star clusters according to our cluster formation model.

One caveat of this model is that bound clouds are not strictly expected to be the sole contributors to star formation: a GMC with essentially ${\it any}$ virial parameter could form some number of stars, in a collapsing sub-region. However, we do expect the overall stellar population to be heavily weighted toward those formed in a bound progenitor cloud, because star formation efficiency is expected to fall off rapidly as a function of virial parameter \citep{padoan:2012.sfe, dale:2017,kim:2021.hii.gmcs.mhd}. In effect, we model the expected continuous-but-steep transition between starless and star-forming clouds with decreasing virial parameter as a step-function at $\alpha_{\rm vir}=2$.



\section{Results}

\label{section:results}
\begin{figure*}
     \centering
     \includegraphics[width=\textwidth]{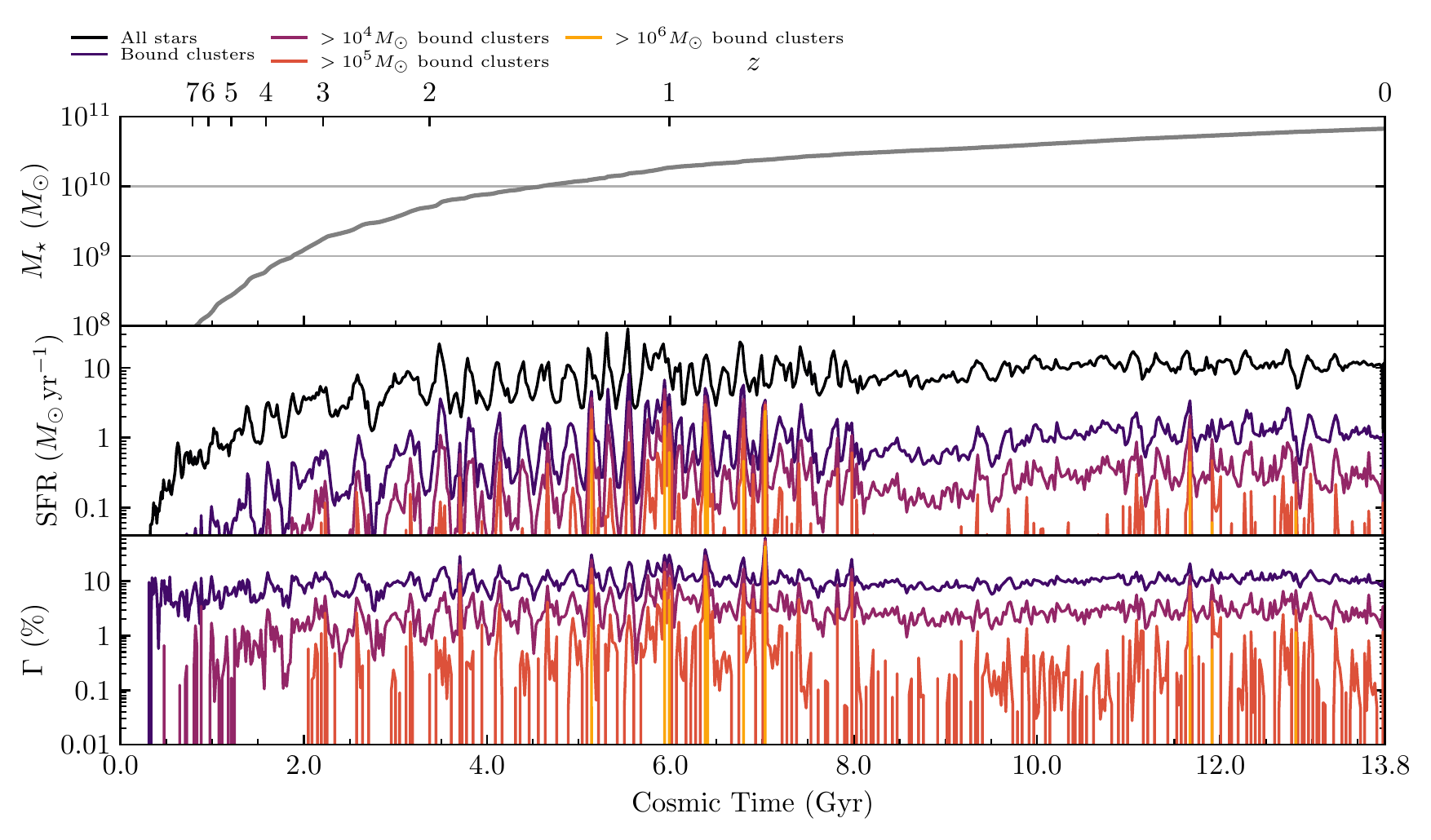}\vspace{-8mm}
     \caption{Star and star cluster formation history of the simulated galaxy. {\it Top:} Total stellar mass of the host galaxy as a function of cosmic time. {\it Middle}: Star formation rate in each simulation snapshot, showing the contributions of bound clusters above different mass cuts. {\it Bottom}: Cluster formation efficiency of the simulated galaxy across cosmic time, for all bound clusters and various cluster mass cuts. $<10^4 M_\odot$ clusters are produced with an efficiency varying only by a factor of $\sim 3$, $>10^4 M_\odot$ clusters near-constantly with an efficiency varying by an order of magnitude, and more massive clusters only episodically.}
     \label{fig:t_vs_gamma}
 \end{figure*}

In Figure \ref{fig:t_vs_gamma} panel 1 we plot the total stellar mass of the galaxy as a function of time. At $z=0$ the galaxy has a total stellar mass of $6.7\times 10^{10}M_\odot$. At $z=0$ this galaxy has some noted differences from the Milky Way. It is not part of a ``Local Group" that contains another comparably massive galaxy within $1\rm \,Mpc$. Its gas fraction is $\sim 20\%$, versus $\sim 10\%$ for the Milky Way, and  its star formation rate at the present epoch is significantly higher, $\sim 10 M_\odot \rm yr^{-1}$, compared to the observed $\sim 2 M_\odot \rm yr^{-1}$ \citep{licquia:2015.milkyway.sfr,blandhawthorn:2016.milkyway}.
As such, this galaxy is later-forming than the Milky Way, having a higher SFR at late times and a lower SFR at early times, giving it roughly equal stellar mass at $z=0$. This fact will prove important when we interpret the age and metallicity statistics of the massive clusters formed in the model, vis-a-vis those found in the Milky Way (\S \ref{sec:azr}).  




 \subsection{Cluster formation efficiency}
  \begin{figure*}
     \centering
     \includegraphics[width=\textwidth]{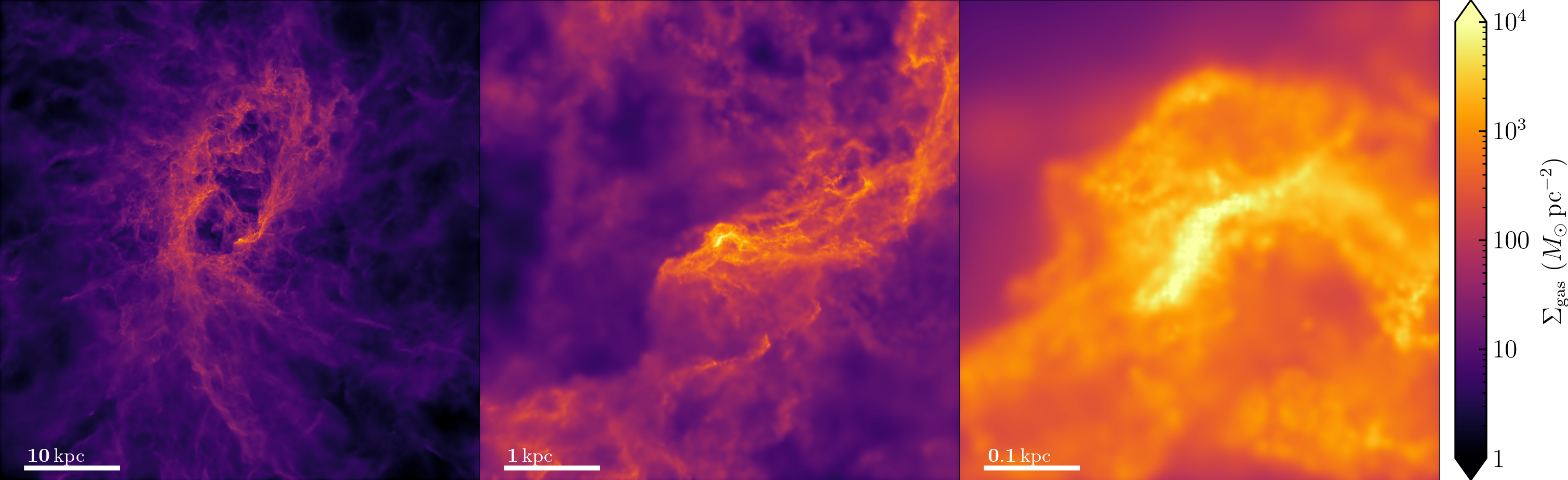}
     \vspace{-5mm}
     \caption{Gas surface density at the formation site of the most massive ($7\times 10^6 M_\odot$) cluster formed in the history of the simulated galaxy, in a $3\times 10^7 M_\odot$ cloud with mean surface density $\Sigma_{\rm GMC} \sim 1200 M_\odot \rm pc^{-2}$ at $z \sim 0.8$. The cloud is found at the edge of a large bubble or cavity, and the galaxy still has a highly irregular morphology. 3D animations of this cloud and the 2 next-most-massive cluster progenitor clouds can be viewed \href{http://www.starforge.space/gbof_movie.mp4}{here}.}
     \label{fig:allrender}
 \end{figure*}

The galactic cluster formation efficiency $\Gamma$ varies in time and space according to local GMC properties in the simulation according to the scalings given \ref{eq:fbound} and the sampling procedure described in \S\ref{section:methods}. Overall, this approach finds that 13\% of all stars formed in this galaxy are in bound clusters following the dispersal of their natal cloud -- in other words, the vast majority of stars are never members of clusters that remain bound after gas expulsion.
 In Figure \ref{fig:t_vs_gamma} panels 2 and 3 we break down the star formation rate and cluster formation efficiency $\Gamma$ for different mass ranges of bound clusters. On average, $\Gamma \sim 10\%$, with no clear systematic trend with cosmic time. But over $\lesssim 100 \rm Myr$ timescales, $\Gamma$ can undergo significant swings, between $\sim 1-100\%$. From comparison of panels 2 and 3 it is evident that these swings follow modulations in the star formation rate of the galaxy, indicating that variations in star formation activity are driving variations in GMC properties (and hence $\Gamma$ in turn). However, $\Gamma$ is clearly not a one-to-one function of SFR, as the most intense starbursts do not necessarily have the most efficient cluster formation -- rather, we will show that $\Gamma$ depends more sensitively on the {\it intensity} of star formation $\Sigma_{\rm SFR}$ than the total SFR, as has been inferred from observations \citep{hollyhead:2016.ngc1566.clusters}.

 To illustrate how GMC properties can vary dramatically from those typically observed in present-day nearby disk galaxies \citep[e.g.][]{bolatto:2008.gmc.properties}, giving rise to high cluster formation efficiencies, Figure \ref{fig:allrender} plots the surface density of gas in the galaxy surrounding the most prodigious cluster-forming cloud in our catalogue, during the large spike in $\Gamma$ at $z \sim 0.8$ evident in Fig \ref{fig:t_vs_gamma}. This cloud is found at the edge of a ``superbubble", a large cavity evacuated by a major stellar feedback event, similar to some of the more extreme cases noted in high-redshift galaxies simulated in \cite{ma_2020_fire_cluster_formation}. The cloud has a mass of $ 3\times 10^7 M_\odot$ and a mean surface density $\Sigma_{\rm GMC} = M_{\rm GMC}/\left(\uppi R_{\rm GMC}^2 \right) = 1200 \,M_\odot\,\rm pc^{-2}$. According to Eqs. \ref{eq:sfe}-\ref{eq:fbound}, this surface density gives the cloud a star formation efficiency of $27\%$ and a cluster formation efficiency of almost unity, allowing it to form the most massive cluster in the history of the galaxy, with a mass of $7 \times 10^6 M_\odot$ and an initial half-mass radius of $5 \rm pc$\footnote{This cloud also produced the ``{\small behemoth}" cluster originally described and studied in \citet{behemoth}.}.
 Born near the galactic center, the cluster has a dynamical friction time much less than a Hubble time, so its mostly likely fate is to spiral into the galactic center and merge into the nuclear star cluster (\citealt{capuzzo:2008.cluster.df,pfeffer:2018.emosaics}; Rodriguez et al. in prep.).

 \begin{figure*}
     \centering
     \includegraphics{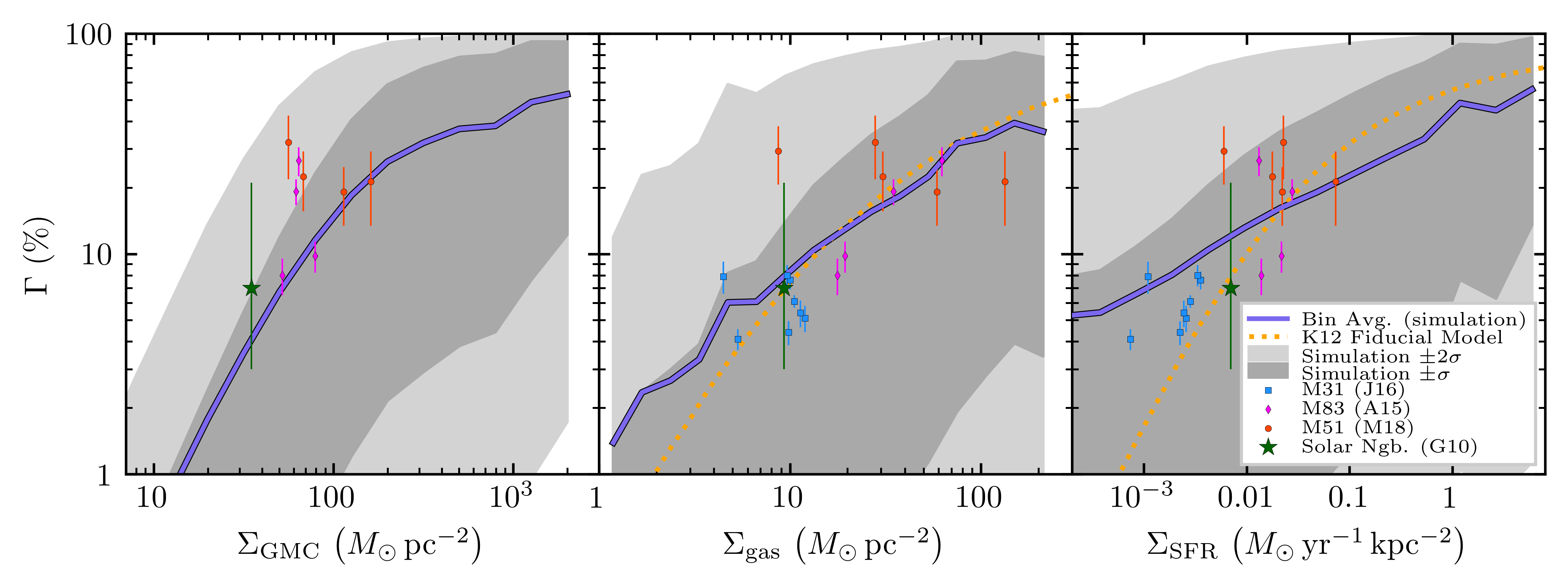}\vspace{-5mm}
     \caption{Model-predicted cluster formation efficiency $\Gamma = \sum M_{\rm cl}/\sum M_{\rm \star}$ as a function of GMC-scale surface density $\Sigma_{\rm GMC}$ (left), kpc-scale galaxy gas surface density $\Sigma_{\rm gas}$ (middle), and kpc-scale star formation surface density $\Sigma_{\rm SFR}$ (right). Solid curves plot the average efficiency ($\sum M_{\rm cl}/\sum M_{\rm \star}$ in a given bin), shaded regions plot the cluster formation efficiency of the clouds that the $16-84\%$ and $5-95\%$ percentiles of stars formed in. We compare with the fiducial model of \citealt{kruijssen:2012.cluster.formation.efficiency} (K12) and several local measurements in M31 \citep{johnson:2016.cluster.formation.efficiency}, M83 \citep{adamo:2015.m83.clusters}, M51 \citep{messa:2018.m51}, and the Solar neighbourhood \citep{goddard:2010.cfe} (see \S\ref{sec:localgamma} for details).
     }
     \label{fig:stuff_vs_gamma}
 \end{figure*}


 \subsubsection{Environmental scaling relations}
 \label{sec:localgamma}
To analyze the relation between the local galactic environment and $\Gamma$, we break down the cloud catalogue in terms of GMC surface density $\Sigma_{\rm GMC}$, local gas density $\Sigma_{\rm gas}$ measured on $1\rm kpc$ scales, and local star formation surface density $\Sigma_{\rm SFR}$, also measured on $1\rm kpc$ scales. Note that only $\Sigma_{\rm GMC}$ is a direct input for our model. To compute $\Sigma_{\rm gas}$ in the vicinity of a cloud, we count the total gas mass within a $1\rm kpc$ radius of the cloud and take $\Sigma_{\rm gas} = M_{\rm gas}/\uppi/\left(1 \rm kpc\right)^2$. We compute $\Sigma_{\rm SFR}$ similarly, estimating the total SFR within $1 \rm kpc$ of each cloud by counting the total stellar mass $< 10 \rm Myr$ old and taking $ \rm SFR \approx M_{\rm star}/10\rm Myr$, and then let $\Sigma_{\rm SFR} = \rm SFR /\uppi/\left(1 \rm kpc\right)^2$.

In Figure \ref{fig:stuff_vs_gamma} we plot the average $\Gamma  = \sum M_{\rm cl} / \sum M_{\rm \star}$ in different bins of $\Sigma_{\rm GMC}$, $\Sigma_{\rm gas}$, and $\Sigma_{\rm SFR}$, and compare these results with various observations and the predictions of the fiducial version of the \citet{kruijssen:2012.cluster.formation.efficiency} (hereafter \citetalias{kruijssen:2012.cluster.formation.efficiency}) analytic model. We plot $\Gamma$ measurements in resolved subregions of M83 \citep{adamo:2015.m83.clusters}, M31 \citep{johnson:2016.cluster.formation.efficiency}, and M51 \citep{messa:2018.m51.2}, using $\Sigma_{\rm gas}$ and $\Sigma_{\rm SFR}$ values provided in those respective works. We use $\Gamma$ values measured with age cuts of $>10 \rm Myr$ and $\lesssim 100 \rm Myr$ from these works,  we also plot the measurement for the Solar neighborhood given in \citet{goddard:2010.cfe}, and use $\Sigma_{\rm gas} = 10 M_\odot \rm pc^{-2}$ and $\Sigma_{\rm SFR} = 7\times 10^{-3} M_\odot \, \rm yr^{-1} kpc^{-2}$ \citep{bovy:2017.solar.ngb}. For $\Sigma_{\rm GMC}$, we use the mass-weighted median values of clouds in the \citet{colombo:2014.m51.gmcs} and \citet{freeman:2017.m83.gmcs} catalogues, in the same respective radial bins as $\Gamma$ was measured, in M51 and M83 respectively, and for the solar neighbourhood we use the fiducial value of $35 M_\odot \, \rm pc^{-2}$ given in  \citet{lada:2020.sigma.gmc}. 

Figure \ref{fig:stuff_vs_gamma} shows that $\Gamma$ exhibits a clear scaling with $\Sigma_{\rm GMC}$, $\Sigma_{\rm gas}$, and $\Sigma_{\rm SFR}$. The correlation
with $\Sigma_{\rm GMC}$ follows directly from the cluster formation model via the dependence of the cloud-scale $f_{\rm bound}$ in Eq. \ref{eq:fbound}, which is physically a consequence of the higher star formation efficiency of denser clouds, Eq. \ref{eq:sfe}.
The $\Sigma_{\rm gas} - \Gamma$ relation agrees well with the fiducial \citet{kruijssen:2012.cluster.formation.efficiency} relation, and has a similar level of agreement with the observations. The $\Sigma_{\rm SFR} - \Gamma$ relation also agrees well with the fiducial \citet{kruijssen:2012.cluster.formation.efficiency} model for $\Sigma_{\rm SFR} > 10^{-2} M_\odot \, \rm yr^{-1} \,kpc^{-2}$, however it predicts a systematically greater $\Gamma$ at lower $\Sigma_{\rm SFR}$, and as a result matches the \citet{johnson:2016.cluster.formation.efficiency} M31 measurements better. This discrepancy with the fiducial \citetalias{kruijssen:2012.cluster.formation.efficiency} model was noted in \citet{johnson:2016.cluster.formation.efficiency} and a modification to the model was proposed that reproduces the observations with similar success.

The most glaring discrepancies with {\it both} our and \citetalias{kruijssen:2012.cluster.formation.efficiency}'s predictions is M51: taken at face value, none of the measurements provided by \citet{messa:2018.m51} (red points) substantiate a systematic trend in $\Gamma$ with {\it any} environmental property considered here. One possible explanation is that the measurements do not fully capture variations in kpc-scale environmental properties: \citet{messa:2018.m51} measured $\Gamma$ in radial bins, but M51 is the prototype for strong spiral structure -- within a given radial bin, $\Sigma_{\rm GMC}$, $\Sigma_{\rm gas}$, and $\Sigma_{\rm SFR}$ can vary systematically as a function of azimuth. Within either our or \citetalias{kruijssen:2012.cluster.formation.efficiency}'s models, cluster formation in a given bin would likely be dominated by the high-density spiral arms, and this could obscure any signal of small $\Gamma$ values expected in the inter-arm regions. 

Another complication of the measurements in M51 is that \citet{messa:2018.m51.2} found fairly steep cluster age distributions in certain regions, suggesting that cluster destruction may reduce the measured value of $\Gamma$ in the $10-100 \rm Myr$ age window significantly. This would not be as much of an issue in M31 and M83, which have much flatter age distributions \citep{bastian:2012.m83.clusters, johnson:2016.cluster.formation.efficiency}.

The average $\Sigma_{\rm gas}$-$\Gamma$ relation is well approximated by the fit
 \begin{equation}
     \Gamma = \min \left( 0.063 \left(\frac{\Sigma_{\rm gas}}{10 M_\odot \,  \rm pc^{-2}}\right)^{0.8}, 1 \right),
 \end{equation}
 and the dependence on $\Sigma_{\rm SFR}$ is approximated by
 \begin{equation}
     \Gamma = \min \left(0.12 \left(\frac{\Sigma_{\rm SFR}}{10^{-2} M_\odot \,  \rm yr^{-1} \, kpc^{-2}}\right)^{0.3},1\right),
     \label{eq:sigma_SFR_vs_gamma}
 \end{equation}
 which is quite similar to the $\Gamma \propto \Sigma_{\rm SFR}^{0.24}$ fit to compiled observational data by \citet{goddard:2010.cfe}.

\begin{figure}
    \centering
    \includegraphics[width=\columnwidth]{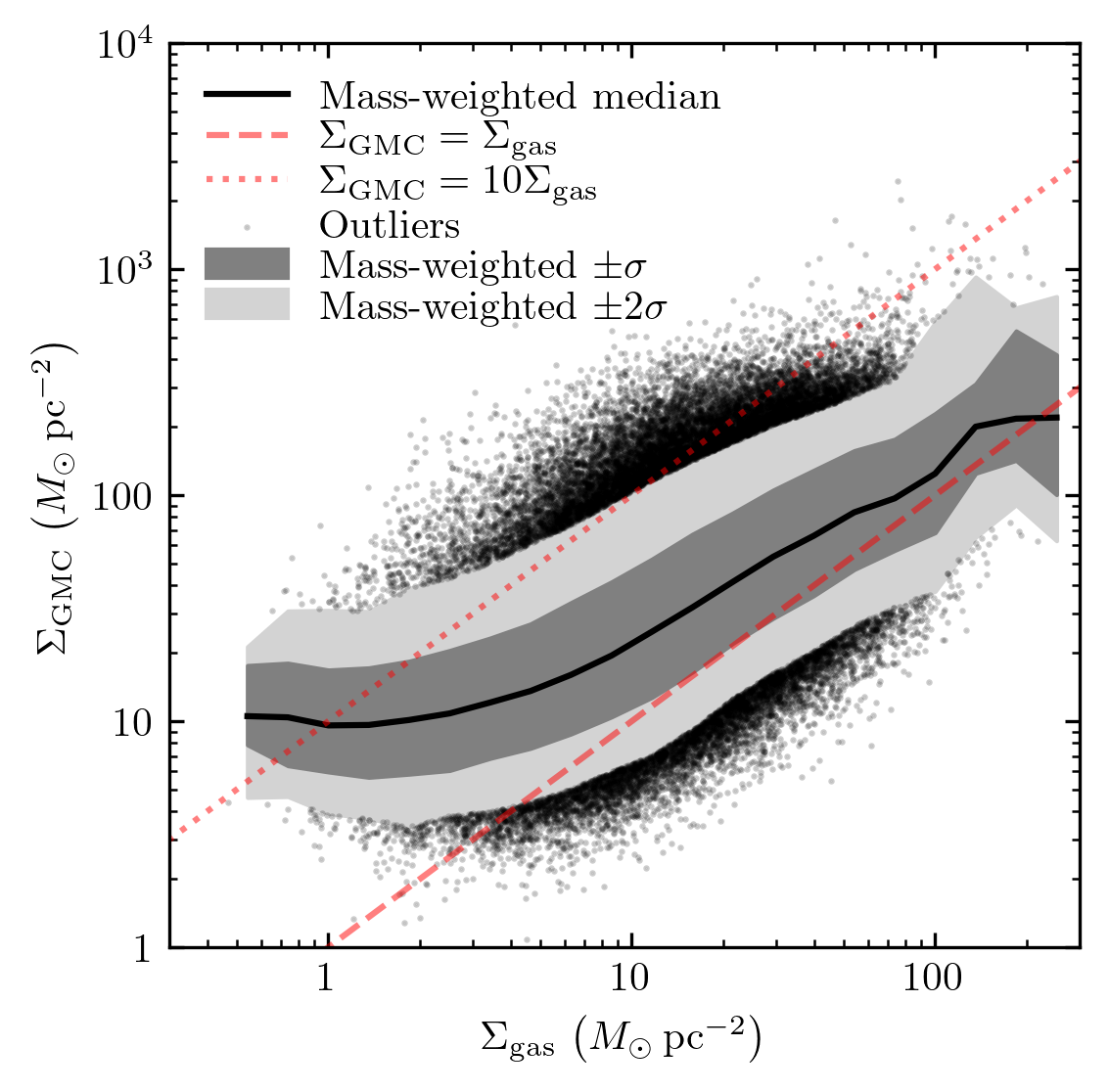}\vspace{-6mm}
    \caption{Relation between the surface density $\Sigma_{\rm GMC}$ of individual bound GMCs in our catalogue, and the average gas surface density $\Sigma_{\rm gas}$ in a 1kpc sphere surrounding each cloud. We plot mass-weighted quantiles binned by $\Sigma_{\rm gas}$, and $>2\sigma$ outliers are plotted as points.
    }
    \label{fig:sigma_gas_v_sigma_GMC} 
\end{figure}

\begin{figure}
    \centering
    \includegraphics[width=\columnwidth]{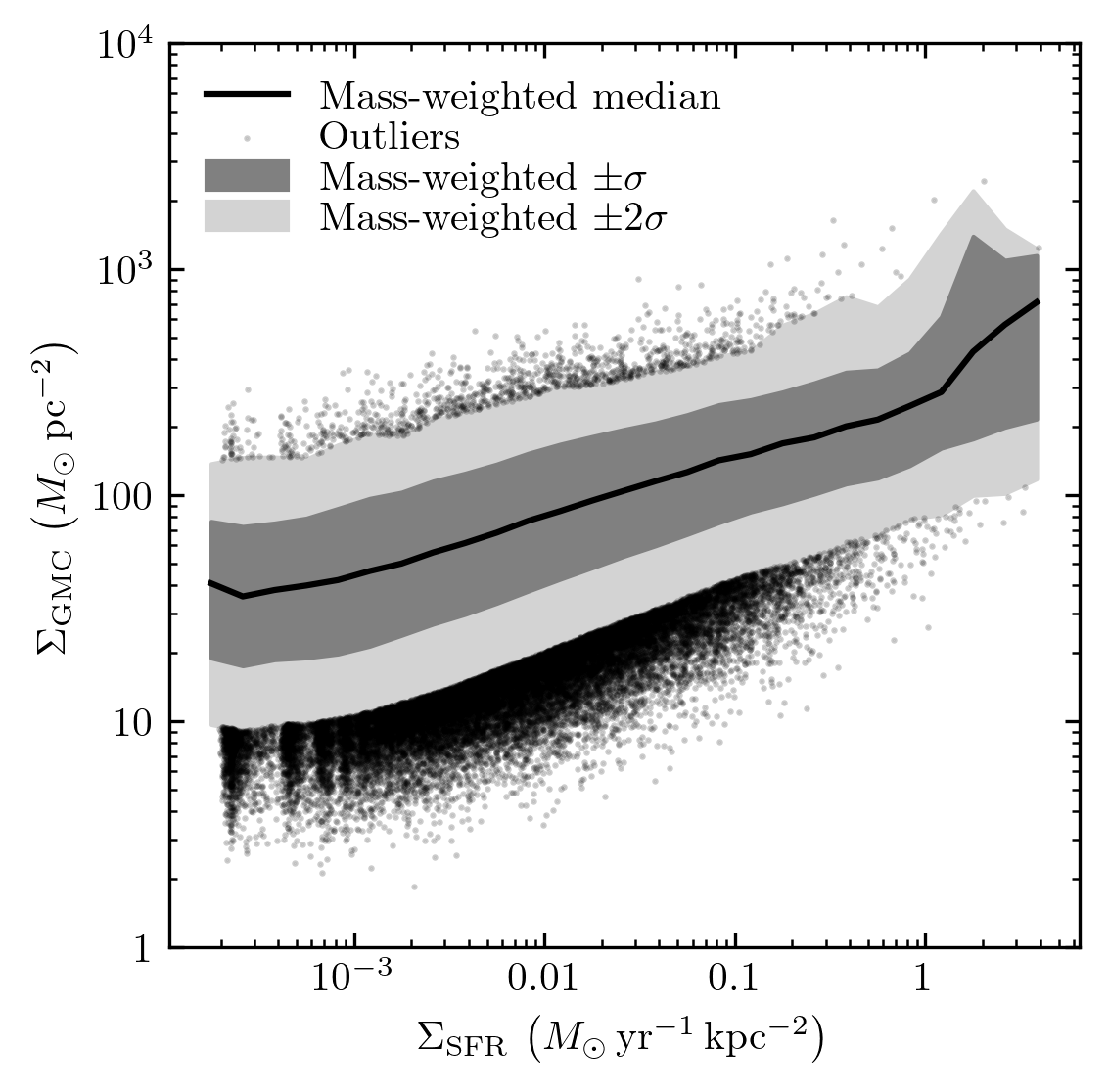}\vspace{-7mm}
    \caption{ Relation between the surface density $\Sigma_{\rm GMC}$ of GMCs in our catalogue, and the average star formation surface density $\Sigma_{\rm SFR}$ in a 1kpc sphere surrounding them.}
    \label{fig:sigma_SFR_v_sigma_GMC} 
\end{figure}

\subsubsection{Relating $\Sigma_{\rm GMC}$ to $\Sigma_{\rm gas}$ and $\Sigma_{\rm SFR}$}
Recalling that $\Sigma_{\rm GMC}$ is the quantity that determines $\Gamma$ within our model, the trends in $\Gamma$ with $\Sigma_{\rm gas}$ and $\Sigma_{\rm SFR}$ require that $\Sigma_{\rm GMC}$ have some systematic scaling with these quantities. We plot these relations in Figures \ref{fig:sigma_gas_v_sigma_GMC} and \ref{fig:sigma_SFR_v_sigma_GMC} - both quantities are similarly predictive of $\Sigma_{\rm GMC}$, with a residual scatter in either relation of $0.6 \rm dex$ across nearly the entire dynamic range. These (mass-weighted) relations and their scatter can be modeled by the fits

\begin{equation}
    \Sigma_{\rm GMC} = \left(8 + 1.8\left(\frac{\Sigma_{\rm gas}}{1 M_\odot \rm pc^{-2}}\right)^{0.9}\right) M_\odot \rm pc^{-2} \pm 0.6 \rm dex,
    \label{eq:sigma_gas_v_sigma_GMC}
\end{equation}
and
\begin{equation}
    \Sigma_{\rm GMC} = \left(14 + \left(\frac{\Sigma_{\rm SFR}}{10^{-2} M_\odot \rm \, yr^{-1}\, kpc^{-2}}\right)^{0.33}\right) M_\odot \rm pc^{-2} \pm 0.6 \rm dex.
\end{equation}
It should be noted that this fit to the $\Sigma_{\rm gas}-\Sigma_{\rm GMC}$ relation is not expected to extrapolate to arbitrarily high $\Sigma_{\rm gas}$, as the asymptotic scaling is $\Sigma_{\rm GMC}\propto \Sigma_{\rm gas}^{0.9}$, implying a crossover point where $\Sigma_{\rm gas} \sim \Sigma_{\rm GMC}$ -- above this point, an average scaling at least as steep as $\propto \Sigma_{\rm gas}$ is necessary, as otherwise the ``clouds" would be voids against the denser environment.

That $\Sigma_{\rm GMC}$ (and the resulting $\Gamma$) should correlate with {\it both} $\Sigma_{\rm gas}$ and $\Sigma_{\rm SFR}$ is unsurprising, as these quantities tend to be highly correlated across a large dynamic range of scales \citep{schmidt:1959,kennicutt:1998.review,heiderman:2010.gmcs,elmegreen:2018.kslaw,pokhrel:2021.gmc.kslaw}.
 
\subsection{Cluster initial mass function}
\label{sec:massfunc}
\begin{figure*}
    \centering
    \includegraphics[width=\textwidth]{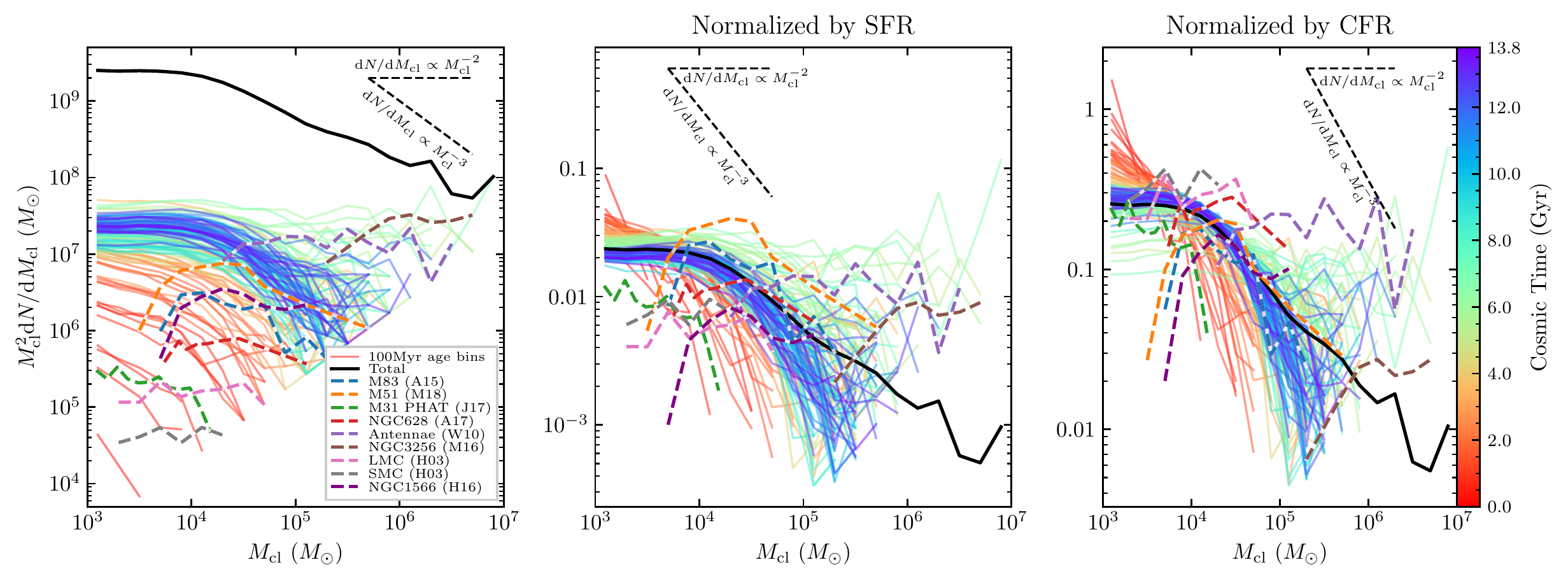}\vspace{-4mm}

    \caption{
    Initial mass functions of bound clusters, plotted as $M_{\rm cl}^2 \mathrm{d}N/\mathrm{d} M_{\rm cl}$, i.e. compensating for the expected dominant $\propto M_{\rm cl}^2$ scaling. We compare the mass function of clusters formed in 100Myr windows across cosmic time (color coded by time) with observed mass functions in nearby galaxies (see \S\ref{sec:massfunc} for data compilation references). Also plotted is the {\it total} mass function integrated across cosmic time (black). {\it Left}: $M_{\rm cl}^2 \mathrm{d}N/\mathrm{d} M_{\rm cl}$ with no additional normalization. {\it Centre}: Like panel 1, but dividing out the total stellar mass formed (bound and unbound) in the respective simulation and galactic age bins (i.e. normalizing by the star formation rate). {\it Right}: Like panel 1, but dividing out the total {\it bound} cluster mass formed in each age bin (i.e. normalizing by the cluster formation rate).
    }
    \label{fig:time_vs_massfunction}
\end{figure*}

\begin{figure}
    \centering
    \includegraphics[width=\columnwidth]{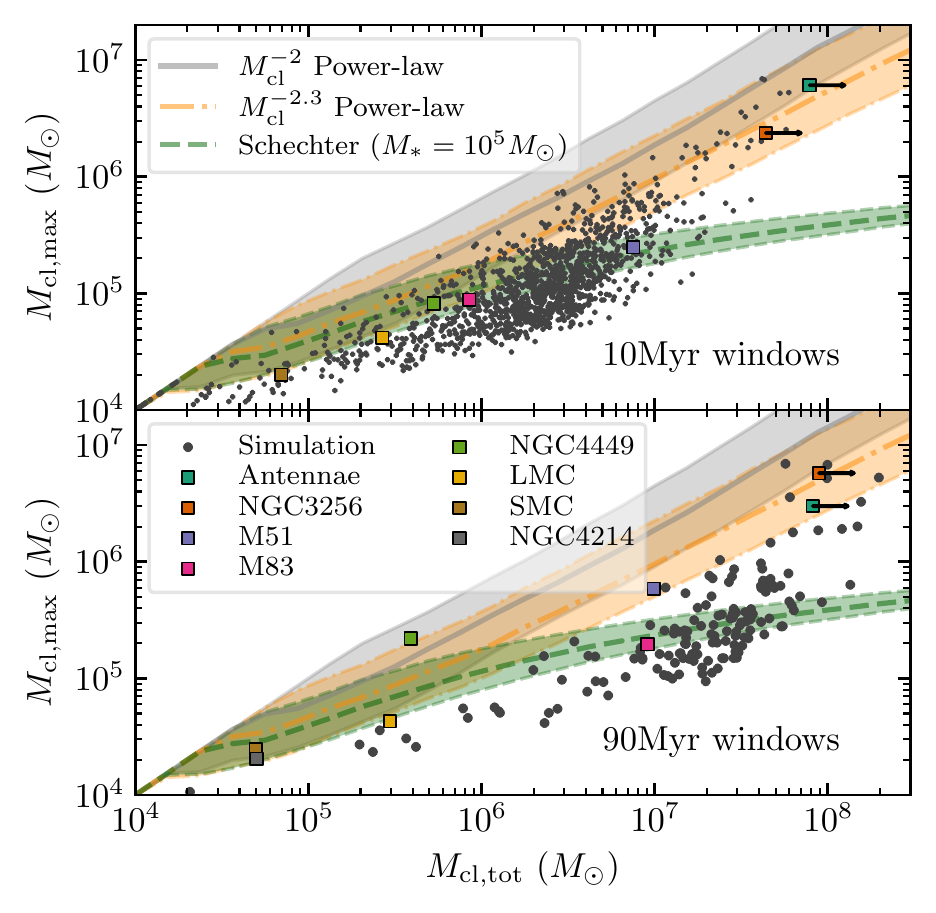}\vspace{-8mm}
    \caption{Relation between the total mass in clusters more massive than $10^4 M_\odot$, and their maximum mass, in 10Myr (top) and 90Myr (bottom) age bins, compared with observations of 1-10Myr old (top) and 10-100Myr old (bottom) clusters in different galaxies, following \citet{mok:2019.clusters}. Points show values across for each $10 \rm Myr$ windows in the simulation, squares are the sample of different galaxies compiled in \citet{mok:2019.clusters}. For comparison we plot the median and $\pm \sigma$ ranges according to three hypotheses for the mass function: a $\propto M_{\rm cl}^{-2}$ or $\propto M_{\rm cl}^{-2.3}$ power-law (here truncated at $10^8 M_\odot$) and a ``Schechter-like" form $\propto M_{\rm cl}^{-2} \exp\left(-M_{\rm cl}/10^5 M_\odot\right)$.}
    \label{fig:mtot_vs_mmax}
\end{figure}

\begin{figure}
    \centering
    \includegraphics[width=\columnwidth]{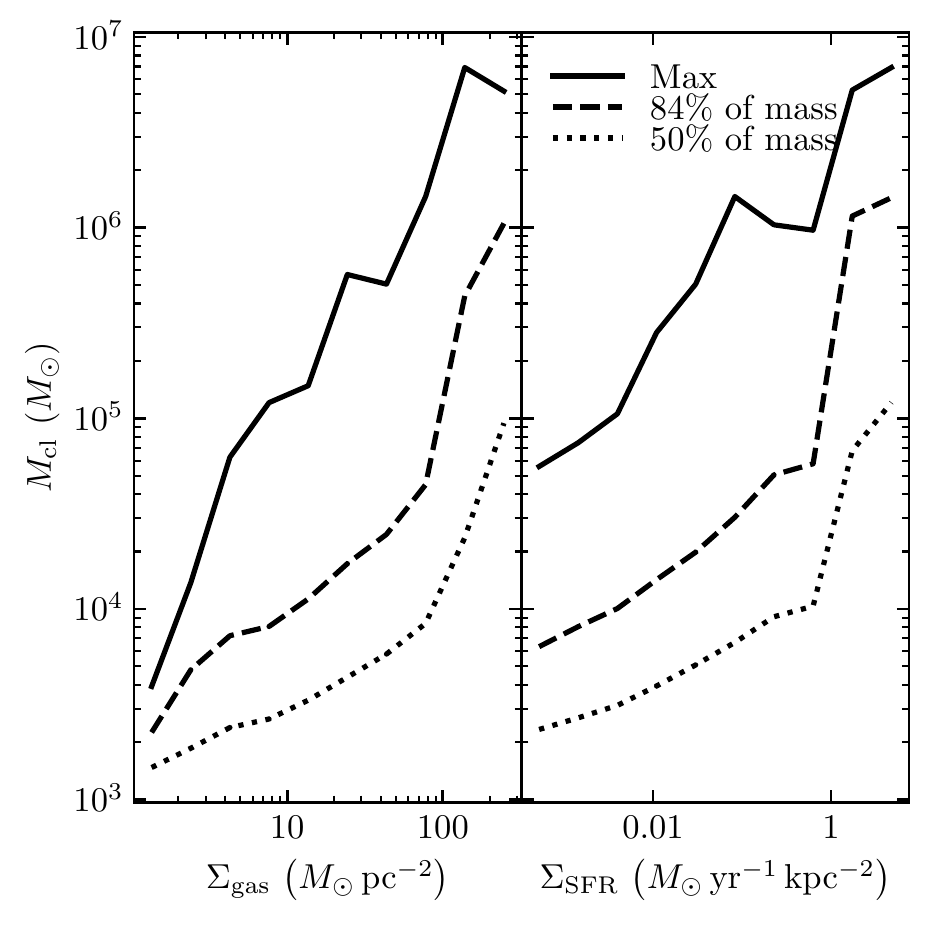}\vspace{-8mm}
    \caption{Mass-weighted percentiles of the cluster mass function (cluster mass below which a given percentage of the total cluster mass lies) binned by $\Sigma_{\rm gas}$ (left) and $\Sigma_{\rm SFR}$ (right), taken over all of cosmic time in the simulation.
    }
    \label{fig:sigmas_vs_massfunc}
\end{figure}
\begin{figure}
    \centering
    \includegraphics[width=0.5\textwidth]{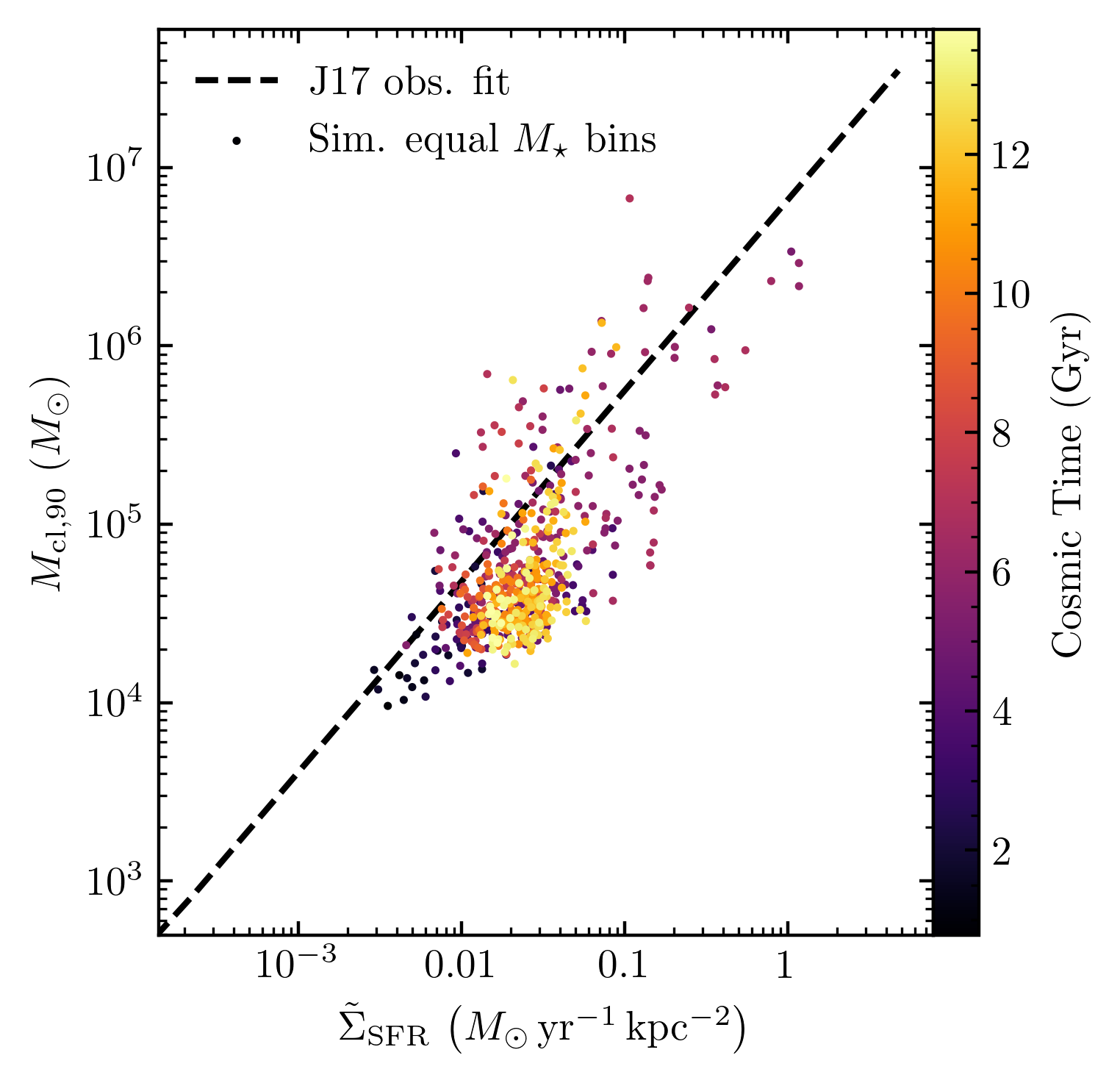}\vspace{-8mm}
    \caption{Relation between the 90th mass percentile of the cluster mass function and the median $\Sigma_{\rm SFR}$ that a star formed in, in time windows from the simulation during which equal stellar mass forms. For comparison we plot the fit to measurements of the Schechter cutoff of cluster mass funtions proposed in \citet{johnson:2017.m31.massfunction} (dashed).}
    \label{fig:sigma_SFR_vs_mmax}
\end{figure}


We now examine the initial mass function of the bound clusters formed in our model, recalling that our model samples cluster masses from a local mass function within each GMC, so the integrated galactic mass function will be the result of stacking samples from the variable mass functions of each cloud within a certain age bin. Figure \ref{fig:time_vs_massfunction} plots the mass functions of clusters formed in different $100 \rm Myr$ windows across cosmic time, and the total mass function. We compare these with a variety of mass functions observed in nearby galaxies, generally for clusters in the age range $10-100 \rm Myr$ where possible. These data include catatlogues from the LMC and SMC (H03; \citealt{hunter:2003.lmc.smc.clusters}), M83 (A15; \citealt{adamo:2015.m83.clusters}), M51 (M18; \citealt{messa:2018.m51.2}), the M31 PHAT field (J17; \citealt{johnson:2017.m31.massfunction}), the Antennae (W10; \citealt{whitmore:2010.antennae.clusters}), NGC1566 (H16; \citealt{hollyhead:2016.ngc1566.clusters}), NGC3256 (M16; \citealt{mulia:2016.ngc3256.clusters}), and NGC628 (A17; \citealt{adamo:2017.cfe}).

The clusters formed in the simulation span essentially the entire observed mass range of young star clusters found in nearby galaxies (up to $7 \times 10^6 M_\odot$), with the exception of NGC 7252, which hosts the most massive known young cluster \citep{maraston:ngc7252.w3, bastian:2013.ngc7252.clusters}.
The integrated mass function over cosmic time is fairly bottom-heavy, resembling a power-law $\mathrm{d}N/\mathrm{d} M_{\rm cl} \propto M_{\rm cl}^{-2.5}$. The explanation for this bottom-heavy mass function can be discerned from the diverse mass functions seen at different periods in the galaxy's history: the galaxy form clusters as massive as $\sim 10^7 M_\odot$ only during a couple exceptional episodes, and spends most of its time forming clusters significantly less massive, putting most of the overall bound cluster mass in lower-mass clusters. 

The sequence of mass functions exhibits a a discernible evolution over cosmic time. At early times ($< 3 \rm Gyr$), fewer, lower-mass clusters generally form, but as we reach $\sim 6 \rm Gyr$ ($z \sim 1$) the galaxy experiences its most intense epsiodes of cluster formation, forming clusters as massive as $7 \times 10^6 M_\odot$ (as illustrated in Figure \ref{fig:allrender}). And finally, as we approach $z \sim 0$ the formation of clusters $>10^6 M_\odot$ becomes rarer, and the maximum young cluster mass is typically on the order of $10^5 M_\odot$, as found in various nearby disk galaxies \citep[e.g.][]{adamo:2015.m83.clusters,messa:2018.m51.2}. When plotting the mass function in equal time windows, a large portion of the variation is simply driven by variations in the overall star and star cluster formation rate -- to control for this, Figure \ref{fig:time_vs_massfunction} panels 2 and 3 plot the mass functions controlling for the total stellar mass and total cluster mass formed in the respective time windows. This collapses most of the variation, but even when controlling for the total formation rate, true variations in the shape of the mass function exist -- the different mass functions tend to vary in slope at the high-mass end, being steeper (or having lower ``truncation" mass) when lower-mass clusters form and shallowest when the highest-mass clusters form.

When analyzing the shape of cluster mass functions, it is important to control for the total mass of clusters in the sample, as a poorly-sampled mass function with a large or nonexistent cutoff can be difficult to distinguish from a mass function with a genuine cutoff \citep{mok:2019.clusters}. In Figure \ref{fig:mtot_vs_mmax} we distinguish between different hypotheses for the mass function by plotting how the mass of the most massive cluster varies as a function of the total cluster mass above $10^4 M_\odot$, for both 10 and 90 Myr time windows in the simulation, compared to observed 1-10Myr and 10-100Myr old cluster populations respectively.
For comparison we plot the expected scalings assuming various different forms for the overall mass function - a pure power-law $\mathrm{d}N/\mathrm{d} M_{\rm cl} \propto M_{\rm cl}^{-2}$, a slightly steeper $\mathrm{d}N/\mathrm{d} M_{\rm cl} \propto M_{\rm cl}^{-2.3}$, and a Schechter-like form with a cutoff of $10^5 M_\odot$, $\propto M_{\rm cl}^{-2} \exp\left(-M_{\rm cl}/10^5 M_\odot\right)$. The data -- in both the simulations and observations -- do not conform perfectly to any one assumed form of the mass function. Rather, they appear to span a sequence that agrees well with the Schechter-like form when the total mass is lower, and then break from this pattern toward a regime that agrees better with the $M_{\rm cl}^{-2.3}$ power-law form. From this is is clear that our mass functions defy a description in terms of any one simple, time-invariant power-law or Schechter-like form. The cluster mass function varies intrinsically across environment and cosmic time.

\subsubsection{Environmental dependence of the mass function}

In \S\ref{sec:localgamma} we found that variations in cluster formation efficiency can be traced to environmental variations in GMC properties, so naturally this is also the case for the mass function, explaining the variations along the sequence of points plotted in Figure \ref{fig:mtot_vs_mmax}. In Figure \ref{fig:sigmas_vs_massfunc} we plot the mass-weighted quantiles of the cluster mass function as a function of $\Sigma_{\rm gas}$ and $\Sigma_{\rm SFR}$: within each bin: the cluster mass below which a certain percent of the total cluster mass in each bin lies. We find that the mass scale of clusters increases monotonically with both environmental properties considered. This trend is primarily driven by the increase in star and star cluster formation efficiency in the denser GMCs found in denser environments (cf. Figs \ref{fig:sigma_gas_v_sigma_GMC},\ref{fig:sigma_SFR_v_sigma_GMC}), rather than an increase in the mass scale of GMCs. For example, the most massive $7 \times 10^6 M_\odot$ cluster formed in a $3 \times 10^7 M_\odot$ cloud (Fig. \ref{fig:allrender}) with high efficiency due it its high $\gtrsim 10^3 \mspc$ mean surface density, whereas the most massive cloud in the cloud catalogue is $2\times 10^8 M_\odot$ but had a mean surface density of $\sim 100 \mspc$, so its most massive cluster was only $10^5 M_\odot$.

\citet{johnson:2017.m31.massfunction} proposed a similar correlation between the cutoff of the mass function and the average value of $\Sigma_{\rm SFR}$ in a galaxy, fitting a power-law relation $M_{\ast} \propto \langle \Sigma_{\rm SFR} \rangle^{1.1}$ to mass function fits from M31, M83, M51, and the Antennae. Direct comparison to this result is complicated by the fact that not all of our mass functions {\it are} well fit by a Schechter-like model with a constrained cutoff, but in Figure \ref{fig:sigma_SFR_vs_mmax} we plot the mass below which 90\% of the total cluster mass exists, in time windows containing equal formed stellar mass, as a function of $\tilde{\Sigma}_{\rm SFR}$, the median $\Sigma_{\rm SFR}$ that a star formed in in each time window. This has a similar relation to that found in \citet{johnson:2017.m31.massfunction}.

\subsection{Initial mass-radius relation}
\label{sec:massradius}
\begin{figure}
    \centering
    \includegraphics[width=\columnwidth]{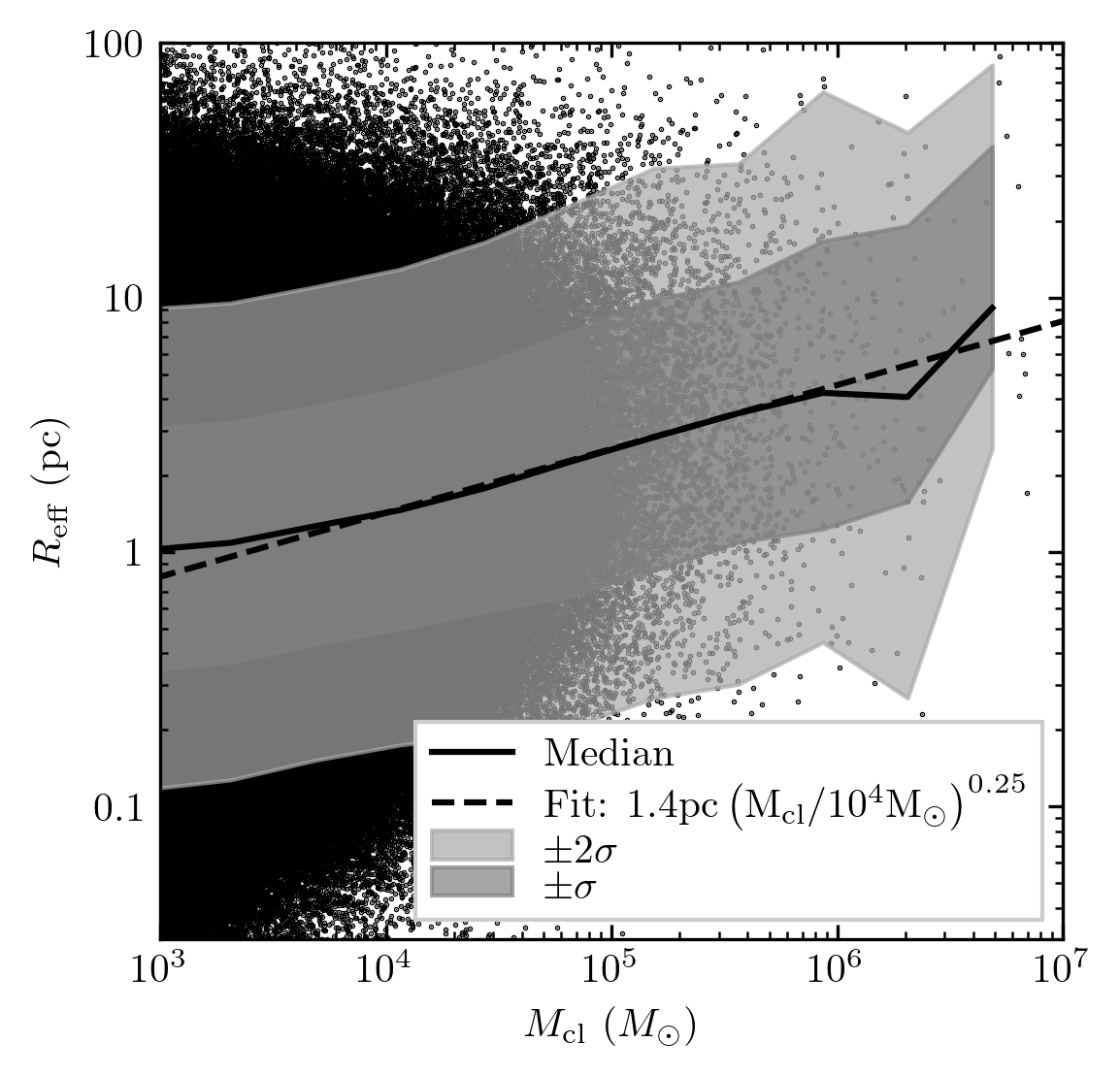}\vspace{-8mm}
    \caption{Mass-radius relation of the entire model cluster population, plotting the projected half-mass radius $R_{\rm eff}$. Overlaid are number-weighted median, $\pm \sigma$, and $\pm 2\sigma$ quantiles in different mass bins, and an unweighted least-squares power-law fit giving $R_{\rm eff} \propto M_{\rm cl}^{0.25}$.}
    \label{fig:mass_radius_relation}
\end{figure}

In Figure~\ref{fig:mass_radius_relation} we plot the 2D projected half-mass radii $R_{\rm eff}$ of the simulated star clusters as a function of their initial mass $M_{\rm cl}$. The initial mass-radius relation has large scatter ($\sim 0.5 \rm dex$),  which is nearly independent of cluster mass. This scatter is the result of convolving the {\it intrinsic} scatter of $\pm 0.4 \rm dex$ set by cluster formation physics resolved in the \citetalias{grudic:2020.cluster.formation} simulations with the properties of the GMC catalogue, which introduce additional scatter because the median cluster size scales $\propto \Sigma_{\rm GMC}^{-1}$. Fitting the entire dataset to a power-law gives 
\begin{equation}
    R_{\rm eff} = 1.4 \rm pc \left(\frac{M_{\rm cl}}{10^4 M_\odot}\right)^{0.25},
    \label{eq:reff_fit}
\end{equation}
i.e. slightly shallower than a constant-density relation $R_{\rm eff} \propto M_{\rm cl}^{1/3}$, and in agreement with the slope measured from the aggregated LEGUS catalogue of star clusters in nearby star-forming galaxies \citep{brown:2021.cluster.mass.radius}. 

The normalization of Eq \ref{eq:reff_fit} is $\sim 40\%$ smaller than the relation fitted in \citet{brown:2021.cluster.mass.radius}, and our scatter is roughly twice as great.Similar discrepancies were noted in \citet{grudic:2020.cluster.formation} when comparing the cluster radii predicted by this model with the measurements by \citet{ryon:2015.m83.clusters}, and we discussed several possible explanations. First, the numerical simulations from which the cluster formation model is derived are  subject to significant uncertainties because they are sensitive to uncertain assumptions about the unresolved conversion of gas to stars \citet{ma_2020_fire_cluster_formation,hislop:2022.cluster.formation}. Some discrepancy in the predicted stellar phase-space density is therefore expected. But it should also be noted that we are predicting only {\it initial} cluster radii, and stellar and dynamical evolution will tend to increase the size of the cluster over time. And the scatter is also expected to decrease as the cluster population evolves, because while dense clusters expand to fill their tidal Roche lobe, under-dense clusters will be stripped of their outer parts through tidal shocks, or destroyed entirely \citep{gieles.renaud:2016.cluster.evol}. 

\begin{figure}
    \centering
    \includegraphics[width=\columnwidth]{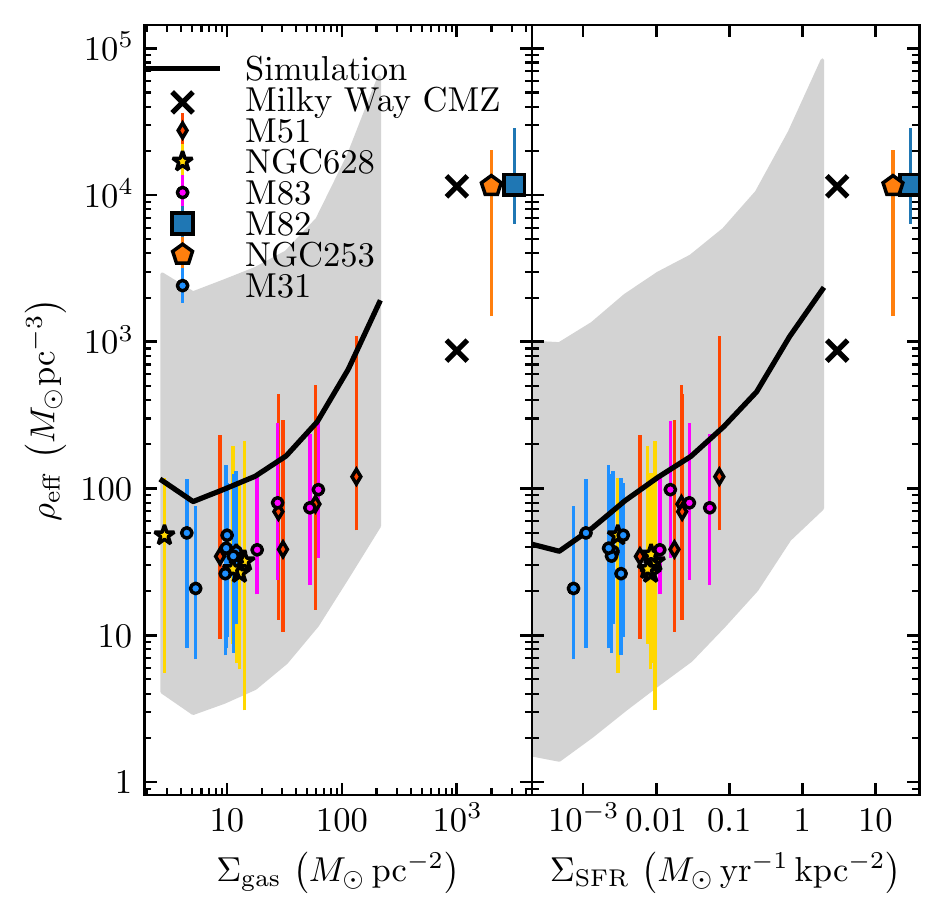}\vspace{-8mm}
    \caption{Scaling of 3D cluster density with $\Sigma_{\rm gas}$ (left) and $\Sigma_{\rm SFR}$ (right). The line and shaded interval plot the number-weighted median and $\pm \sigma$ quantiles in different bins of $\Sigma_{\rm gas}$ and $\Sigma_{\rm SFR}$, which we compare with observations in different regions of various galaxies (see \S\ref{sec:massradius} for details).}
    \label{fig:rhoeff_vs_env}
\end{figure}

We have also examined the cluster size-mass relations when controlling for their natal environmental conditions $\Sigma_{\rm gas}$ and $\Sigma_{\rm SFR}$ -- when binning the data by these quantities, we generally find a best-fit mass-radius relation consistent with $R_{\rm eff} \propto M_{\rm cl}^{1/3}$ \citep[e.g.][]{fall:2012.universal.cluster.formation}, but the proportionality factor varies with environment.
Hence, the mass-radius relation can be described by an environmentally-varying 3D density, with intrinsic scatter, and the shallower relation of the aggregate sample emerging because more massive clusters tend to form in denser GMCs, and hence tend to be smaller. 
In Figure \ref{fig:rhoeff_vs_env} we plot the number-weighted median and $16-84\%$ range of the 3D half-mass density $\rho_{\rm eff} = 3M_{\rm cl}/\left(8 \uppi r_{\rm eff}^3\right)$, where $r_{\rm eff}$ is the 3D half-mass radius, as a function of $\Sigma_{\rm gas}$ and $\Sigma_{\rm SFR}$, compared with various observations. Cluster sizes and masses in different subregions of M83 are taken from \citet{ryon:2015.m83.clusters}, and $\Sigma_{\rm gas}$ and $\Sigma_{\rm SFR}$ from \citet{adamo:2015.m83.clusters}. Sizes, masses, and environmental properties in different M31 PHAT fields are taken from \citet{johnson:2015.phat.clusters,johnson:2017.m31.massfunction, johnson:2016.cluster.formation.efficiency} respectively. Cluster masses in M51 and NGC628 are taken from the data compilation and density profile fits performed by \citet{brown:2021.cluster.mass.radius} on data from the LEGUS survey \citep{calzetti:2015.legus,adamo:2017.cfe,cook:2019.legus}, and radially-binned $\Sigma_{\rm gas}$ and $\Sigma_{\rm SFR}$ from \citet{messa:2018.m51} and \citet{chevance:2020.gmcs} for M51 and NGC628 respectively. Lastly we use data for M82, NGC253, and the Milky Way central molecular zone (CMZ) compiled by \citet{choksi:2019.cluster.mass.radius}.

Figure \ref{fig:rhoeff_vs_env} shows that the median cluster density scales systematically with both $\Sigma_{\rm gas}$ and $\Sigma_{\rm SFR}$ in our model, and we find $\pm 1.1\rm dex$ of residual scatter about the median. A similar trend in cluster density is also found in the observational data, as was noted by \citet{choksi:2019.cluster.mass.radius}. Our model consistently overpredicts the median density of clusters compared to the observed clusters, but we note that our model predicts {\it initial} cluster densities, while the observations are of $\sim 1-100 \rm Myr$ old clusters. These clusters have had time to lose mass and expand under the influence of stellar evolution and dynamical evolution, so we expect observed evolved clusters to be less dense than the predicted initial density. The {\it scatter} in initial density is considerably greater than the scatter in density of observed clusters, but again, we expect that evolutionary processes will tend to reduce the scatter in the densities of a cluster population: clusters that are initially ``too small" will tend to puff up due to internal evolution, while clusters that are initially ``too large" are more susceptible to stripping, shocking, and destruction in the galactic environment. It will be possible to examine this hypothesis by modeling the dynamical evolution of each cluster (Rodriguez et al., in prep.).

\subsection{Age-metallicity relation}
\label{sec:azr}
Finally, we analyze the age-metallicity relation of candidate globular clusters, which we take to be clusters more massive than $10^5 M_\odot$ for our present purposes. The relation between the ages and metallicities of ancient star clusters in a galaxy contains information about the galaxy's formation history: each cluster surviving to the present day provides a snapshot of the metallicity of the environment in which it formed, with an associated timestamp. The cluster metallicity can be related to a certain stellar mass of the host galaxy, via the redshift-dependent mass-metallicity relation \citep[e.g.][]{2004ApJ...613..898T, mannucci_hiz_mzr,2013ApJ...779..102K, ma_2016_fire_azr}, so provided the metallicities of old globular clusters reflect those of their host galaxy as whole, they place constraints upon the stellar mass of the progenitor galaxy at a certain time.

\begin{figure}
    \centering
    \includegraphics[width=\columnwidth]{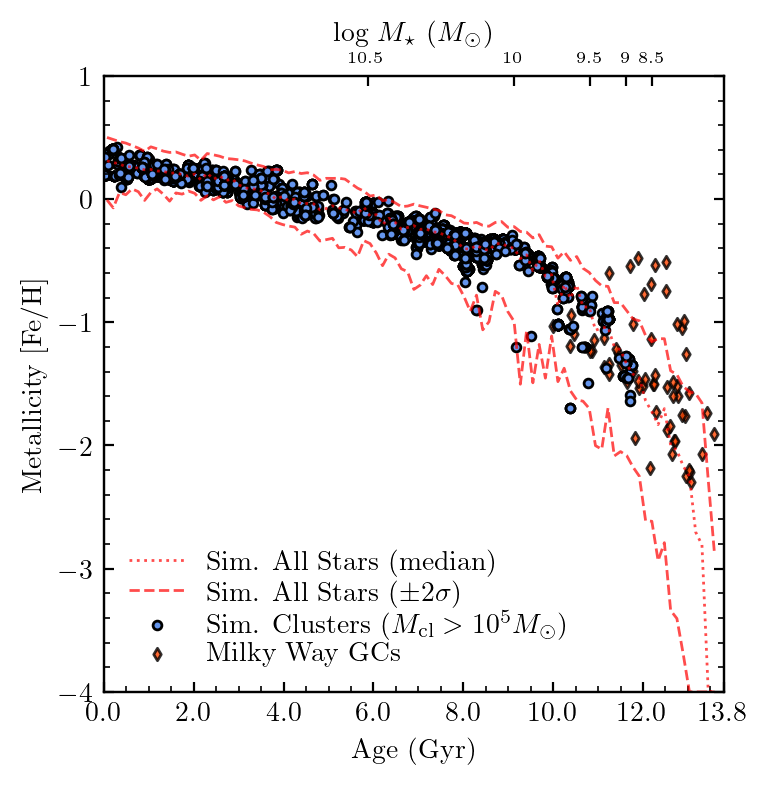}\vspace{-8mm}
    \caption{Age-metallicity relation of $>10^5 M_\odot$ clusters formed in our model (circles), the entire stellar population of the simulation (red lines), and Milky Way globular clusters (diamonds, data compiled by \citealt{kruijssen:2019.emosaics}). On top we mark the simulated main galactic host stellar mass at various times.}
    \label{fig:AZR }
\end{figure}

 Within the model, clusters inherit their abundances from their progenitor cloud in the simulation, and the simulation itself  includes a model for turbulent mixing in the ISM that gives realistic metallicity variations in the galaxy \citep{escala:fire.metallicity, bellardini:fire.metallicity}. The age-metallicity relation of massive clusters in our model, and of stars in the galaxy as a whole, are plotted in Figure \ref{fig:AZR }, which we compare with data for Milky Way globular clusters compiled in \citet{kruijssen:2019.emosaics.mw}. Our main result is that {\it massive clusters do not form with a metallicity substantially different from other stars forming within the galaxy}, i.e. massive cluster formation is not strongly biased toward more or less metal-rich regions. Hence, if globular clusters formed as a result of the normal star formation process at high redshift \footnote{Normal galactic star formation need not account for {\it all} globular clusters: more-exotic mechanisms of extragalactic globular cluster formation have been proposed, e.g. \citet{peebles:1968.gcs.recombination} and more recently \citet{naoz:2014.sigos}.}, their age-metallicity statistics would trace the overall properties of the progenitor galaxies faithfully.  

Figure \ref{fig:AZR } also suggests that our simulated cluster population is not a good model for the globular cluster population of the Milky Way: the first $>10^5 M_\odot$ cluster forms at $\sim 2 \,\rm Gyr$ ($z \sim 3$), and at this time a significant number of globular clusters should have already formed, in the Milky Way and in other galaxies \citep{beasley:2000.gc.ages,woodley:2010.gc.ages,vandenberg:2013.gc.ages,usher:2019.gc.ages}. Moreover, even if massive clusters formed sooner, the age-metallicity relation in the model cannot reproduce the sequence of old, red (metal-rich) globular clusters, which are believed to be the population that formed in-situ in the Milky Way \citep{forbes.bridges:2010.gcs,kruijssen:2020.kraken}. The galaxy would have to be significantly more enriched at early times to host an old, red population.

 
\section{Discussion}
\label{section:discussion}

\subsection{Does cluster formation efficiency vary with environment?}
\label{section:dosgammavary}

Many observational works have argued that cluster formation efficiency $\Gamma$ does vary with galactic environment \citep{bastian:2008.cfe,goddard:2010.cfe,adamo:2015.m83.clusters,johnson:2016.cluster.formation.efficiency}, and many ensuing theoretical works have found that this is to be expected from the physics of star formation \citep{kruijssen:2012.cluster.formation.efficiency,li:2017.cluster.formation,pfeffer:2018.emosaics, lahen_2019_cluster_formation}. On the other hand, \citet{chandar:2017.cfe} argued that some or all of the scaling in $\Gamma$ that other works inferred could be explained by contamination of the cluster sample by young ($\lesssim 10 \rm Myr$), unbound systems, calling the scaling of $\Gamma$ into some question.

In \S\ref{sec:localgamma} we found that denser (higher $\Sigma_{\rm gas}$) and more actively star-forming (higher $\Sigma_{\rm SFR}$) regions host systematically denser self-gravitating GMCs (with higher $\Sigma_{\rm GMC}$, see Figures \ref{fig:sigma_gas_v_sigma_GMC},\ref{fig:sigma_SFR_v_sigma_GMC}). In turn, our GMC-scale model predicts that the denser GMCs in these regions form stars more efficiently, resulting in higher $\Gamma$ in that region.
Hence, we concur with the growing consensus of theoretical predictions of variable $\Gamma$, and have put it on firmer footing using simulations with a self-consistent GMC population formed from cosmological initial conditions. With that said, we do concur with \citet{chandar:2017.cfe} that reliable estimates of $\Gamma$ without stellar kinematic information are only possible for cluster age ranges that 1) are too old to not be gravitationally bound and 2) are too young to have experienced significant mass loss and disruption, and caution against the over-interpretation of $\Gamma$ measurements from cluster populations that may not satisfy these criteria \citep[e.g.][]{adamo:2020.hipeec.clusters}.  

More generally, given the modern understanding of the star cluster formation process, it is increasingly difficult to imagine a scenario wherein $\Gamma$ does {\it not} vary with environment: $\Gamma$ has been extensively shown to correlate with the local SFE of the host GMC, in analytic theory \citep{hills:1980,mathieu:1983}, idealized stellar dynamics calculations modeling gas removal \citep{tutukov:1978,lada:1984,baumgardt.kroupa:2007,smith:2011.cluster.assembly,smith:2013.cluster.assembly}, and hydrodynamics simulations with spatially-resolved star and star cluster formation and gas removal by stellar feedback \citep{li:2019.cfe,lahen_2019_cluster_formation,grudic:2020.cluster.formation}. Star formation efficiency, in turn, has been predicted to vary with GMC properties, a prediction that follows from a very general considerations of limiting cases of momentum- and energy-conserving feedback \citep{fall:2010.sf.eff.vs.surfacedensity,krumholz:2018.star.cluster.review}, which has been almost unanimously supported by GMC simulations that treat stellar feedback and simulate a range of GMC properties \citep{hopkins:fb.ism.prop,dale:2014,grudic:2016.sfe,geen:2017, howard:2017, kim:2018,kim:2021.hii.gmcs.mhd,fukushima:2021.gmc.rhd.sim}. \footnote{Note that the agreement of different simulations on this issue is only {\it qualitative} at present -- the SFE predicted for a given GMC model still varies widely between simulation suites, in part due to the variety of prescriptions in use for unresolved star formation and feedback, their chief uncertainty \citep{elephant}.} The missing link up to this point has been the relation between GMC properties and the $\sim \rm kpc$-scale quantities $\Sigma_{\rm gas}$ and $\Sigma_{\rm SFR}$, which has not been possible to study in nearby galaxies in a homogeneous fashion. But in this work we have shown that GMC properties are coupled to environmental properties, so $\Gamma$ follows in turn.

\subsection{Comparison with previous cosmological star cluster formation studies}
\subsubsection{E-MOSAICS}
E-MOSAICS \citep{pfeffer:2018.emosaics} is a suite of simulations coupling semi-analytic cluster formation and evolution prescriptions to the EAGLE cosmological hydrodynamics simulations \citep{2015MNRAS.446..521S}. Unlike the FIRE-2 simulation used in the present work, E-MOSAICS simulations do not explicitly resolve GMCs and the multi-phase interstellar medium, relying instead on a sub-grid ``effective equation of state" prescription to model the dynamics of the ISM \citep{springel:2003.subgrid.ism}, and using the \citet{reinacampos:2017.cluster.model} prescription to model the GMC population according to coarse-grained (~kpc-scale) ISM properties. The GMC mass function is then mapped onto the cluster mass function assuming a constant SFE of 10\% and a cluster formation efficiency derived from a local formulation of the \citep{kruijssen:2012.cluster.formation.efficiency} model. These simulations also model the ongoing evolution of clusters on-the-fly in the simulation, accounting for stellar evolution and a variety internal and external dynamical processes, which we have not attempted here (see however Rodriguez et al., in prep., in which we model cluster evolution in post-processing). This makes it possible to comment on the age distribution of young clusters (which is affected by mass loss and disruption), as well as the population of clusters surviving to $z=0$, in a large sample of simulated galaxies.

Our model and the \citet{kruijssen:2012.cluster.formation.efficiency} model agree fairly well on the environmental dependence of $\Gamma$ (Figure \ref{fig:stuff_vs_gamma}), so the prescription used in E-MOSAICS appears to be a reasonably good approximation of our findings derived from explicitly-resolved ISM structures. However, the assumption of constant SFE does not agree with the consensus of numerical simulations with stellar feedback (see references in \S\ref{section:dosgammavary}), including the \citetalias{grudic:2020.cluster.formation} cluster formation model we have used here, in which SFE scales as a function of $\Sigma_{\rm GMC}$. The assumed constant value of $10\%$ may be a reasonable average value weighted by stellar mass formed, but we expect it to vary with environment, given the environmental variations in $\Sigma_{\rm GMC}$ we find here. For example, the cloud shown in Figure \ref{fig:allrender} has an overall SFE of $27\%$ according to our model. This may weight massive cluster formation more heavily toward regions of denser ISM.

Inspection of the GMC population modeled in E-MOSAICS according to the prescription of \citet{reinacampos:2017.cluster.model} also reveals some discrepancies with the GMC population found in FIRE simulations by \citet{guszejnov_GMC_cosmic_evol}. Their model hypothesizes that the largest possible collapsing gas mass is on the order of the Toomre mass, and when this is applied to the E-MOSAICS simulations it predicts the existence of self-gravitating clouds in excess of $10^{11}M_\odot$ (see \citealt{pfeffer:2018.emosaics} Fig. 5). In comparison, the most massive self-gravitating gas structure formed self-consistently in the \citet{guszejnov_GMC_cosmic_evol} catalogue we have used is $2 \times 10^8 M_\odot$. Even if this mass is to be identified only with the ``collapsed fraction" that is identified as the cluster progenitor cloud in the model, E-MOSAICS simulations host numerous clouds with $M_{\rm GMC} > 10^{10} M_\odot$. \citet{li:2020.smuggle.gmcs} pointed out that the properties of GMCs formed in FIRE-like simulations with resolved ISM structure can be still somewhat sensitive to adopted sub-grid feedback and/or star formation prescriptions, so we do not necessarily consider the \citet{guszejnov_GMC_cosmic_evol} cloud properties definitive, but the mass function variation seen in \citet{li:2020.smuggle.gmcs} was not at the level needed to explain a cloud mass discrepancy of 2 orders of magnitude. Thus, at present there appears to be a disconnect between the semianalytic theory of GMC mass functions applied to the EAGLE simulations, and what is found in numerical simulations with explicit ISM structure.

E-MOSAICS simulations assumed a constant initial cluster radius, surveying various values $r_{\rm eff} = 1.5-6 \rm pc$ and adopting a fidicial value of $4 \rm pc$. As noted in \citet{choksi:2019.cluster.mass.radius} and the present work (\S\ref{sec:massradius}), more recently-available data show evidence of a variable mass-radius relation, taking the form $R_{\rm eff}\propto M_{\rm cl}^{1/3}$, with a varying proportionality factor (e.g. Fig. \ref{fig:mass_radius_relation}). Adopting such a relation would make low-mass clusters smaller and high-mass clusters larger, which would affect their susceptibility to the tidal environment in turn. However, because the relation is shallow we expect the scaling relation itself to have modest effects, as shown by \citet{pfeffer:2018.emosaics}. Likely more important is the significant {\it scatter} found in simulations and observations: this could significantly broaden the range of cluster sizes, and the resulting range of possible dynamical histories.

Lastly, the E-MOSAICS simulations have been used to predict and interpret the age-metallicity relation of globular clusters, with a large sample size of Milky Way-mass galaxies \citep{kruijssen:2019.emosaics,kruijssen:2019.emosaics.mw}. These works do find galaxies that fill the region of age-metallicity space occupied by the Milky Way's globular clusters (c.f. Figure \ref{fig:AZR }), but this appears to lie at the upper envelope of the range spanned by the different simulations -- simulations that form massive clusters relatively late like ours appear to be common in their sample as well.


\subsubsection{Li et al. ART simulations}

In a series of studies, \citealt{li:2017.cluster.formation,li:2018.cluster.zooms,li:2019.cluster.zooms} performed a suite of cosmological zoom-in simulations of Milky Way-mass galaxy progenitors, run with the ART adaptive mesh refinement code \citep{kravtsov:art,agertz:2013.new.stellar.fb.model, semenov:2016.sgs.turb}. Like ours, their simulations did marginally resolve the multi-phase ISM, with a spatial resolution of $6 \rm pc$, so they were able to model the formation of individual GMCs, and cluster formation in turn, modeling cluster formation as a process of accretion and feedback with various subgrid physics prescriptions.

Qualitatively, all of the conclusions reached in these works concerning cluster formation efficiency and the initial mass function of star clusters agree with ours: denser galactic environments produce more top-heavy mass distributions of clusters, with higher efficiency. In particular, \citet{li:2017.cluster.formation} correlate the mass function and cluster efficiency with {\it merger} activity specifically, with mergers leading to more efficient cluster formation. Quantitatively, the predictions of the Li et al. simulations depend very sensitively upon the assumed sub-grid star formation efficiency \citep{li:2018.cluster.zooms}, with lower subgrid efficiencies resulting in lower cluster masses. No $\Sigma_{\rm SFR}$-$\Gamma$ relation presented in that work matches ours especially well in all environments, as the relation is generally shallow compared to ours.

Although the  cluster initial mass function found in \citet{li:2017.cluster.formation} tended to be quite Schechter-like with a typical slope of $\sim -2$, the nominally improved \citet{li:2018.cluster.zooms} suite found a relatively steep (slope between -2 and -3) mass function, similar to the mass function typically found in our model (\S\ref{sec:massfunc}). Observed mass function slopes have are typically around  $\sim -2$ (e.g. \citealt{krumholz:2018.star.cluster.review}, Figure 5), but these can be affected by resolution and completeness effects, and in Figure \ref{fig:time_vs_massfunction} we do find fair agreement with the shapes of cluster mass functions derived from catalogs in nearby galaxies \citep[e.g.][]{adamo:2015.m83.clusters,messa:2018.m51.2}. 



\subsubsection{FIRE simulations}
\citet{kim:2018.fire.gcs} and \citet{ma_2020_fire_cluster_formation} used the FIRE and FIRE-2 frameworks respectively to model the formation of bound star clusters {\it on-the-fly} in the simulations at high redshift, in contrast to the post-processed approach explored here. Those simulations arrived at similar conclusions to us regarding the formation mechanism of the most massive clusters: the sites of massive bound cluster formation were found to be very high pressure and/or surface density ($\gtrsim 10^4 M_\odot \rm pc^{-2}$ (similar to e.g. the scenario shown in Figure \ref{fig:allrender}), achieving high star formation efficiency. Notably, these simulations directly demonstrated that it is possible to achieve such conditions at $z \gtrsim 5$, despite the lack of massive clusters forming at that time in the present work. 

In \citet{ma_2020_fire_cluster_formation} in particular we emphasized that the results of this type of simulation were sensitive to the choice of star formation prescription. To further extend the predictive power of cosmological simulations to detailed predictions of cluster properties on-the-fly, a SF prescription that resolves inherent uncertainties about star formation on small scales is needed. Progress on this front may now be possible by comparing with GMC simulations with individual self-consistent star formation simulations that overlap with the GMC masses that are marginally resolvable in galaxy simulations \citep{guszejnov:isothermal.mhd.sf,starforge.methods,2021MNRAS.502.3646G}.

\subsection{Differences from the Milky Way's globular cluster population}
\label{sec:discussion:milkywaygcs}

In \S\ref{sec:azr} we noted important differences between the age and metallicity statistics of the simulated cluster population and the Milky Way globular cluster population: our model produces no $>10^5 M_\odot$ bound clusters in the first 2 Gyr ($z>3$), and does not reproduce the ``red" population of old, metal-enriched globular clusters.  The most obvious explanation for this is that the simulated galaxy's star formation history is so different from that of the Milky Way: as mentioned in \S\ref{section:results}, the simulated galaxy has similar $z=0$ stellar mass but $\sim 5\times$ higher $z \sim 0$ SFR than the Milky Way, in part due to its relatively high gas fraction of 20\% (a common feature in FIRE-2 Milky Way-like galaxies, see \citealt{2020MNRAS.498.3664G}). To attain similar $z=0$ mass this way, its SFR had to be lower than the MW at early times. The mean SFR of the Milky Way progenitor in the first 2 Gyr has been inferred to be $\sim 5 M_\odot \rm yr^{-1}$ \citep{Snaith_2014}, much greater than the mean $1 M_\odot \,\rm yr^{-1}$ in the first 2 Gyr of our simulation (Figures \ref{fig:t_vs_gamma}). If the SFR was as high as the Milky Way, but concentrated in the same area, the average value of $\Sigma_{\rm SFR}$ would be $\sim 5\times$ greater, increasing the cluster formation efficiency by a factor of $1.6$ (Eq. \ref{eq:sigma_SFR_vs_gamma}) and the upper cutoff of the cluster mass function by a factor of $\sim 6$ (e.g. Figure \ref{fig:sigma_SFR_vs_mmax}), allowing massive clusters to form much sooner. Shifting star formation from late to early times would also make the model more Milky Way-like by suppressing the mass scale of the cluster mass function at late times, which typically has a truncation of $\sim 10^5 M_\odot$ at $z \sim 0$ (Fig \ref{fig:time_vs_massfunction}), more massive than the most massive young clusters in the Milky Way (several $10^4 M_\odot$ at most, \citealt{portegies-zwart:2010.starcluster.review}).

\citet{santistevan:2020.fire.m12.sfh} surveyed the star formation histories of Milky Way-mass galaxies in wider FIRE-2 simulation suite, and found that galaxies in Local Group analogues with two Milky Way-mass galaxies in close proximity form preferentially earlier than isolated Milky Way-mass galaxies like the one we have considered here. Thus, environment may be a factor that differentiates the star formation history of the present model from that of the Milky Way. However, the maximum mass achieved by any galaxy at the 2Gyr mark in \citet{santistevan:2020.fire.m12.sfh} was $\sim 4 \times 10^9 M_\odot$, so these other galaxies would still have difficulty achieving the SFR intensity and metallicity needed to reproduce the old, red GC population. However, they note how various constraints suggest that a star formation history more like the one simulated here -- reaching 50\% of the $z=0$ stellar mass at $z\sim 1$ --  may be typical among galaxies with $M_{\rm 200} \sim 10^{12} M_\odot$ \citep[e.g.][]{behroozi:2019.universe.machine}. If so, the galaxy simulated here -- and its cluster population -- may be more representative of a typical galaxy of this mass, and we would expect the Milky Way's GC population to be systematically ($\sim 2-3 \rm Gyr$) older than a typical galaxy of this mass. 

Even if old, massive globular clusters are produced, old {\it red} ($\left[\rm Fe/H\right] \sim -0.5 $) globular clusters may be difficult to obtain within our framework, even if the early star formation was more rapid. Let us assume the redshift-dependent relation between galactic stellar mass and gas metallicity found in \citet{ma_2016_fire_azr}:
\begin{equation}
    \log \left( Z_{\rm gas}/Z_\odot\right)  = 0.35 \log\left(\frac{ M_{\rm \star}}{10^{10} M_\odot}\right) + 0.93 \exp \left(-0.43 z\right) - 1.05,
\end{equation}
and assume that newborn stars and clusters inherent this gas metallicity at a given redshift. Then a hypothetical galaxy that averaged $5 M_\odot \rm yr^{-1}$ in the first $2 \rm Gyr$ would only form clusters with $\left[\rm Fe/H\right] \sim -1$, still half a dex less than the red population. But the redshift dependence predicted for the mass-metallicity relation does depend crucially upon uncertain feedback and stellar physics \citep{2020MNRAS.491.1656A}, so it is possible that the FIRE simulations used to fit the \citet{ma_2016_fire_azr} relation err in the direction of underestimating metal retention.  

Lastly, it is also worth emphasizing here that {\it forming} the clusters is a necessary, but not sufficient condition for obtaining the population at $z=0$ -- once formed, the clusters are subject to mass loss and disruption in the galactic environment. Detailed predictions of the surviving $z=0$ GC population require a treatment of the dynamical evolution of clusters in the galactic environment, which has been performed in other simulation setups \citep{li:2017.cluster.formation, pfeffer:2018.emosaics}, and which we defer to future work for the present model (Rodriguez et al., in prep.).

\section{Conclusions}
\label{section:conclusions}
In this work we have modeled the population of young star clusters forming in a simulated Milky Way-mass galaxy, extending the predictions of \citet{grudic:2020.cluster.formation} for cluster formation in individual GMCs to the population of cluster progenitor clouds that form self-consistently across cosmic time in the simulation \citep{guszejnov_GMC_cosmic_evol}. We used this model to study various aspects of the star cluster formation:
\begin{itemize}
    \item The efficiency of bound cluster formation $\Gamma$ is $13\%$ in the simulated galaxy. The efficiency does not exhibit a clear systematic trend with cosmic time, but can vary over a wide range at different periods of the galaxy's history (Fig. \ref{fig:t_vs_gamma}). Much of this variation is explained by variations in galactic ISM conditions: there is clear relation between $\Gamma$ and the local $\Sigma_{\rm gas}$ and $\Sigma_{\rm SFR}$ (Fig. \ref{fig:stuff_vs_gamma}), as measured on $\sim 1\rm kpc$ scales in the galaxy. This is because these quantities correlate with the surface density of self-gravitating GMCs (Figs \ref{fig:sigma_gas_v_sigma_GMC},\ref{fig:sigma_SFR_v_sigma_GMC}), which determines star and star cluster formation efficiency in turn, according to the \citetalias{grudic:2020.cluster.formation} model. The environmental scalings we found appear to reproduce the successes (and possible failures, e.g. \citealt{messa:2018.m51.2}) of the \citet{kruijssen:2012.cluster.formation.efficiency} model.
    \item The initial mass function of bound star clusters shows significant diversity over different periods of the galaxy's evolution, similar to the range of diversity seen in observations of nearby galaxies (Figure \ref{fig:time_vs_massfunction}). Both the shape and the normalization of the mass function vary intrinsically, and overall the mass function is not described well by any one simple power-law or Schechter-like form (Figure \ref{fig:mtot_vs_mmax}). This sequence of mass function shapes is similar to what is observed in nearby galaxies, and is driven at least in part by environment: denser environments host denser GMCs, which can form stars more efficiently and produce more massive clusters (Figs. \ref{fig:sigmas_vs_massfunc},\ref{fig:sigma_SFR_vs_mmax}).
    \item We find a global, time-integrated size-mass relation for star clusters of $R_{\rm eff} \propto M_{\rm cl}^{0.25}$, similar to the relation inferred from recent star cluster catalogues \citep{choksi:2019.cluster.mass.radius,brown:2021.cluster.mass.radius}. Within a given environment of fixed $\Sigma_{\rm gas}$ or $\Sigma_{\rm SFR}$, the relation is best described by $R_{\rm eff} \propto M_{\rm cl}^{1/3}$, i.e. constant 3D density, but this density varies with environment (Figure \ref{fig:rhoeff_vs_env}), leading to a global relation shallower than $\propto M_{\rm cl}^{1/3}$. Within a given environment we also predict a significant initial {\it scatter} in initial cluster density of $\sim 1.1 \rm dex$. This is larger than what is observed in $\sim 100 \rm Myr$ old cluster populations, suggesting that star formation physics alone cannot explain the size-mass relation: evolutionary processes must be invoked to reduce the scatter.
    \item The age-metallicity relation of massive ($>10^5 M_\odot$) bound star clusters formed in the galaxy is very similar to that of the stellar population as a whole. Our age-metallicity statistics were incompatible of those of Milky Way globular clusters (Figure \ref{fig:AZR }), but it seems plausible that this difference is driven by a difference in star formation histories (\S\ref{sec:discussion:milkywaygcs}), which affect cluster formation via the local scaling relations with e.g. $\Sigma_{\rm SFR}$ we have found.
\end{itemize}
Thus we have been able to study the properties of {\it young} star clusters as they vary across cosmic time and galactic environment. To model populations of {\it evolved} clusters, and old globular clusters in particular, we must extend our model, accounting for stellar evolution, dynamical evolution, and the influence of the surrounding galactic environment \citep[e.g.][]{pfeffer:2018.emosaics}. This will be the subject of our followup work (Rodriguez et al., in prep.).

The major caveat of this work is that both steps of our model -- predicting galactic ISM structure and mapping those structures onto star clusters -- are not yet fully-solved problems. Attempts to do either in a systematic fashion are still relatively new, and invariably rely upon ad-hoc models for unresolved star formation, turbulence, and feedback. As such, we anticipate that detailed predictions of ISM structure and star cluster formation will continue to evolve as the unresolved microphysics of star formation become better understood.

\section*{Acknowledgements}
We thank S. Michael Fall, J. M. Diederik Kruijssen, Angela Adamo, Marta Reina-Campos, Hui Li, Andrey Kravtsov, Nick Gnedin, Gillen Brown, Omid Sameie and Anna Schauer for enlightening discussions that informed and motivated this work. MYG was supported by a CIERA Postdoctoral Fellowship and a NASA Hubble Fellowship (award HST-HF2-51479).  This work was supported by NSF Grant AST-2009916 at Carnegie Mellon University and a New Investigator Research Grant to C.R.~from the Charles E.~Kaufman Foundation.  AW received support from: NSF via CAREER award AST-2045928 and grant AST-2107772; NASA ATP grant 80NSSC20K0513; HST grants AR-15809, GO-15902, GO-16273 from STScI. MBK acknowledges support from NSF CAREER award AST-1752913, NSF grants AST-1910346 and AST-2108962, NASA grant NNX17AG29G, and HST-AR-15006, HST-AR-15809, HST-GO-15658, HST-GO-15901, HST-GO-15902, HST-AR-16159, and HST-GO-16226 from the Space Telescope Science Institute, which is operated by AURA, Inc., under NASA contract NAS5-26555. AL is supported by the Programme National des Hautes Energies and ANR COSMERGE project, grant ANR-20-CE31-001 of the French Agence Nationale de la Recherche. This work used computational resources provided by XSEDE allocation AST-190018 and TACC Frontera allocations AST-20019 and AST-21002. 
CAFG was supported by NSF through grants AST-1715216, AST-2108230,  and CAREER award AST-1652522; by NASA through grant 17-ATP17-0067; by STScI through grant HST-AR-16124.001-A; and by the Research Corporation for Science Advancement through a Cottrell Scholar Award. Figure~\ref{fig:allrender} was generated with the help of FIRE studio \citep{firestudio}, an open source Python visualization package designed with the FIRE simulations in mind.

\section*{Data Availability}

The data supporting the plots within this article are available upon request to the corresponding author. A public version of the {\small GIZMO} code is available at \url{http://www.tapir.caltech.edu/~phopkins/Site/GIZMO.html}. The {\small CloudPhinder} code used to catalogue self-gravitating clouds in the simulation (\ref{sec:catalogue}) is available at \url{http://www.github.com/mikegrudic/CloudPhinder}.


\bibliographystyle{mnras}
\bibliography{master_nodupe} 







\bsp	
\label{lastpage}
\end{document}